\documentstyle[pre,aps,eqsecnum,psfig,multicol]{revtex}

\begin{document}

\draft

\title{Intermittency in Dynamics of
Two-Dimensional Vortex-like Defects}

\author{V. V. Lebedev}

\address{Department of Physics of Complex Systems,
Weizmann Institute of Science, \\
Rehovot 76100, Israel; \\
and Landau Institute for Theoretical Physics, RAS, \\
Kosygina 2, Moscow, 117940, Russia. \\
e-mail: lwlebede@wicc.weizmann.ac.il and lebede@landau.ac.ru}

\date{\today}

\maketitle

\begin{abstract}

We examine high-order dynamical correlations of defects (vortices,
disclinations etc) in thin films starting from the Langevin equation for the
defect motion. We demonstrate that dynamical correlation functions $F_{2n}$ of
vorticity and disclinicity behave as $F_{2n}\sim y^2/r^{4n}$ where $r$ is the
characteristic scale and $y$ is the renormalized fugacity. As a consequence,
below the Berezinskii-Kosterlitz-Thouless transition temperature $F_{2n}$ are
characterized by anomalous scaling exponents. The behavior strongly differs
from the normal law $F_{2n}\sim F_2^n$ occurring for simultaneous correlation
functions, the non-simultaneous correlation functions appear to be much larger.
The phenomenon resembles intermittency in turbulence.

\end{abstract}


\begin{multicols}{2}

\section*{Introduction}

It is well known that defects like quantum vortices, spin vortices,
dislocations and disclinations play an essential role in physics of
low-temperature phases of thin films. Berezinskii \cite{Ber} and then
Kosterlitz and Thouless \cite{KT} recognized that there is a class of phase
transitions in $2d$ systems related to the defects. The main idea of their
approach is that in $2d$ the defects can be treated as point objects
interacting like charged particles. It is usually called Coulomb gas analogy.
The low-temperature phase corresponds to a fluid constituted of bound
uncharged defect-antidefect pairs, which is an insulator, whereas the
high-temperature phase contains free charged particles and can be treated as
plasma. Correspondingly, in the low-temperature phase the correlation length
is infinite whereas in the high-temperature phase it is finite. A huge number
of works is devoted to different aspects of the problem, see, e.g., the
surveys \cite{78KT,80Nel,83Nel,87Min,GG98}. The scheme proposed by Kosterlitz
and Thouless can be applied to superfluid and hexatic films and planar $2d$
magnetics. It admits a generalization for crystalline films, see Refs.
\cite{Solid,88Str}. There are also applications to superconductive materials,
especially to high-$T_c$ superconductors, see, e.g., Ref. \cite{BFGLV}.

The dynamics of the films in the presence of the defects was considered in
the papers \cite{Dynamics1,Dynamics2}. In the works a complete set of
equations is formulated describing both motion of the defects and hydrodynamic
degrees of freedom. Then, to obtain macroscopic dynamic equations, an
averaging over an intermediate scale was performed. At the procedure the
``current density'' related to the defects was substituted by an expression
proportional to the average ``electric field'' and to gradients of the
temperature and of the chemical potential. The resulting equations perfectly
correspond to the problems solved in the works \cite{Dynamics1,Dynamics2}.
Unfortunately, at the procedure an information concerning high-order
correlations of the defect motion is lost. That is the motivation for the
present work where these high-order correlations are examined.

We start from the same ``microscopical'' equations of the defect dynamics as
was accepted in Ref. \cite{Dynamics1}. Following the works we focus mainly on
the case when the motion of the defects is determined by the Langevin equation
describing an interplay between the Coulomb interaction and the thermal
noise. We believe that the approach is correct for hexatic films (membranes,
Langmuir films, freely suspended films). The situation is a bit more
complicated for the vortices in superfluid films because of the Magnus force.
Nevertheless, the equation for the vortices is close to the Langevin
equation, see Ref. \cite{Dynamics1}. Similar equations can be formulated for
the dislocations in crystalline films, see Ref. \cite{Dynamics2}, for the
vortices in superconductors in some interval of scales, see, e.g., Ref.
\cite{BFGLV}, and for the spin vortices in planar $2d$ magnetics. We will not
consider the last cases here, though our scheme is, generally, applicable to
the systems. Treating non-simultaneous correlation functions related to the
defects one should take into account creation and annihilation processes
also. For the purpose we use the Doi technique \cite{76Doi} who demonstrated
that dynamics of classical particles involved into chemical reactions can be
examined in terms of the creation and annihilation operators, like in the
quantum field theory.

We consider correlation functions $F_{2n}$ of the ``charge density'' $\rho$
(vorticity, disclinicity etc) provided that the so-called
renormalized fugacity $y$ is small. The inequality $y\ll1$ is satisfied for
large scales in the low-temperature phase and probably in some region of
scales above $T_c$. In statics, the normal estimate $F_{2n}\sim F_2^n$ is
valid at the condition. Surprisingly, the non-simultaneous high-order
correlation functions $F_{2n}$ appear to be much larger than their normal
estimate $F_2^n$. In the low-temperature phase the phenomenon reveals an
anomalous scaling on large scales. The reason for such unusual behavior is
that the main contribution to high-order non-simultaneous correlation
functions is associated with rare single defect-antidefect pairs. The
situation resembles the intermittency phenomenon in turbulence, see, e.g.,
Ref. \cite{Frish}. It can also be compared with non-trivial tails of
probability distribution functions in the physics of disordered materials, 
see, e.g., Refs. \cite{LGP,efetov}. Some preliminary results were published
in the paper \cite{99Leb}.

\begin{figure}
\centerline{\psfig{figure=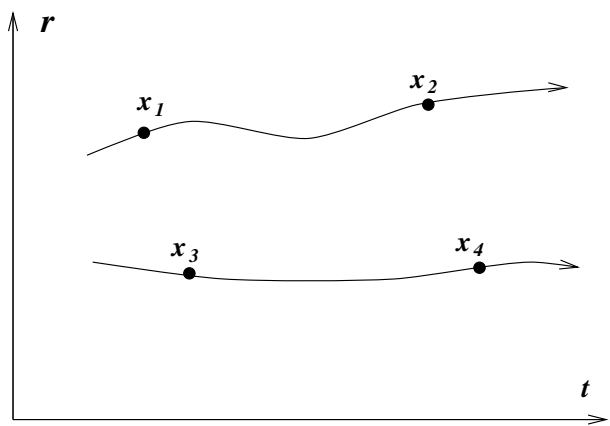,width=8cm}}
\caption{Trajectories passing through given points.}
\label{vort8}
\end{figure}

\begin{figure}
\centerline{\psfig{figure=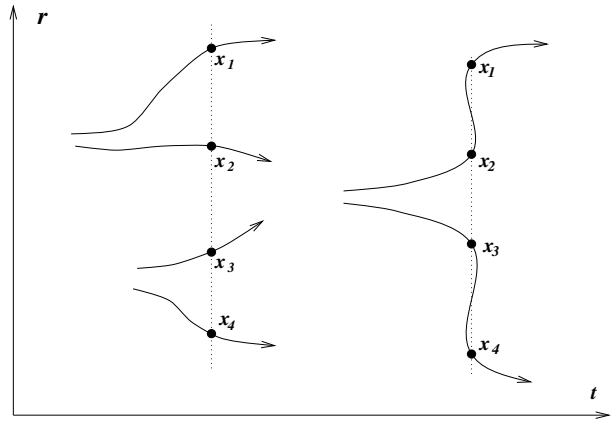,width=8cm}}
\caption{Possible and impossible trajectories
passing through four points at a given time moment.}
\label{vort7}
\end{figure}

Let us give a qualitative explanation of the phenomenon. To obtain a non-zero
contribution to the correlation function
$F_{2n}(t_1,\dots,t_n;{\bbox r}_1,\dots,{\bbox r}_n)$ one must consider
trajectories of the particles passing through the points
${\bbox r}_1, \dots, {\bbox r}_n$ at the time moments $t_1, \dots, t_n$. The
situation is illustrated in Fig. \ref{vort8}. The ``single-pair''
contribution has to be compared with a ``normal'' contribution associated
with a number of defect-antidefect pairs. Though the normal contribution
contains an additional large entropy factor it has also an additional small
factor related to a small probability to observe a defect-antidefect pair
with a separation larger than the core radius. As a result of the
competition, the normal contribution appears to be smaller. To avoid a
misunderstanding, let us stress that the arguments do not work for the
simultaneous correlation functions. The reason is that trajectories of two
defects cannot pass through $n>2$ points simultaneously, see Fig.
\ref{vort7}.

Our paper is organized as follows. In Section \ref{basic} we remind some basic
facts concerning static properties of the $2d$ defects and their dynamics and
then we shortly review the Doi technique \cite{76Doi} suitable for our
problem. In Section \ref{quantum} we develop a diagrammatic representation
for dynamical objects and examine the two-particle conditional probability
which is extensively exploited in the subsequent consideration. In Section
\ref{renorm} we demonstrate how renormalization of different parameters can
be obtained in the framework of our dynamic approach. Actually, the
renormalization is reduced to the well known static renormalization
group equations. In Section \ref{corr} we consider correlation functions 
of the ``charge density'' and ground the properties announced above. In 
Section \ref{helium} we generalize our procedure for the case of superfluid 
films. In the Conclusion we discuss the main results of our work and their 
possible relations to other systems. Some calculations are placed into 
Appendices.

\section{Basic Relations}
\label{basic}

Static properties of the system of the vortex-like defects in thin films can
be described quite universally. The starting point of the description is the
free energy associated with the defects
\begin{equation}
{\cal F}=-\sum_{i\neq j}{T\beta}\,n_in_j
\ln\left(\frac{|{\bbox x}_i-{\bbox x}_j|}{a}\right)+\sum_j \mu(n_j) \,,
\label{H04} \end{equation}
where the subscripts $i,j$ label defects, ${\bbox x}_i$ are positions of the
defects, $a$ is a cutoff parameter of the order of the size of the defect core,
$n_i$ are integer numbers determining the ``strength'' of the defects, $\beta$
is a dimensionless $T$-dependent factor and $\mu$ is the energy associated with
the core. The expression (\ref{H04}) is correct for quantum vortices in
superfluid films, for disclinations in hexatic films, and for spin vortices in
$2d$ planar magnets. For dislocations in crystalline films the expression
(\ref{H04}) has to be slightly modified \cite{Solid}, but the main peculiarity
of the free energy, the logarithmic dependence on the separation, remains the
same.

The Gibbs distribution $\exp(-{\cal F}/T)$ corresponding to the energy
(\ref{H04}) can be treated as the partition function of two-dimensional point
particles with charges $n_j$, $\beta$ playing a role of the ``inverse
temperature''. The parameter $\beta$ can be considered also as the Coulomb
coupling constant. Basing on the electrostatic analogy one can introduce the
``charge density''
\begin{equation}
\rho({\bbox r})=\sum\limits_j n_j
\delta\left({\bbox r}-{\bbox x}_j\right) \,.
\label{den} \end{equation}
The quantity $\rho$ is vorticity for superfluid films and disclinicity for
hexatic films. We will treat the case when defects are produced by thermal
fluctuations. Since both creation and annihilation processes conserve the
``charge'' we should accept that the total charge is zero:  $\sum_j n_j=0$.
It leads to the constraint
\begin{eqnarray} &&
\int{\rm d}^2 r\,\rho({\bbox r})=0 \,,
\label{Ch1} \end{eqnarray}
where the integration is performed over the total area of the specimen.

Below we assume that for $|n|>1$ the core energy $\mu(n)$ is so large that
such defects are hardly created. Then only defects with the charges
$n_i=\pm1$ should be taken into account. We will call the objects with the
charges $n_i=1$ defects and the objects with the charges $n_i=-1$
antidefects. Because of the constraint $\sum_j n_j=0$ there can be
simultaneously $N$ defects and $N$ antidefects in the system. Thus, the
partition function of the system can be characterized via a set of
probability distribution functions ${\cal P}_{2N}$ depending on coordinates
of $2N$ ``particles''. In accordance with Eq.  (\ref{H04}) the functions can
be written as
\begin{eqnarray} &&
{\cal P}_{2N}({\bbox x}_1, \dots, {\bbox x}_{2N})
\nonumber \\ &&
=Z^{-1}\left(\frac{y_0}{a^2}\right)^{2N}
\exp\left\{\sum_{i\neq j}{\beta}\,n_in_j
\ln\left(\frac{|{\bbox x}_i-{\bbox x}_j|}{a}\right)\right\} \,,
\label{parti} \end{eqnarray}
where $Z$ is the sum over states and the quantity $y_0=\exp(-\mu/T)$ is
usually called fugacity. The possibility to neglect charges with $|n|>1$
implies that the fugacity is small.

The low-temperature (insulator) phase can be treated as a system constituted of
bound defect-antidefect pairs. In the high-temperature (plasma) phase there are
unbound charges which essentially influence the system on scales larger than
the correlation length $r_c$. We will treat the low-temperature phase and the
region of scales between $a$ and $r_c$ in the high-temperature phase where one
can neglect the role of the unbound charges and only the bound
defect-antidefect pairs have to be taken into account. The presence of the
pairs in the system leads to non-trivial ``dielectric'' properties of the
medium. As a result the interaction between the charges is modified, the effect
can be described in terms of a scale-dependent ``dielectric constant'' of the
medium as is suggested in Ref. \cite{KT}. In other words, the effective 
coupling constant $\beta$ becomes dependent on the separation between the 
charges.

The scale dependence of $\beta$ can be described in the framework of the scheme
proposed by Kosterlitz, \cite{74Kos}. Namely, the partition function
of the system can be integrated over separations of the defect-antidefect pairs
between the core size ${a}$ and a scale $r$. After the procedure, that can be
interpreted as shifting the core radius $a\to r$, the form of the probability
distribution functions (\ref{parti}) is reproduced (with $r$ instead of $a$),
but the parameters $\beta$ and $y$ are renormalized. The $r$-dependence of
$\beta$ and $y$ is determined by the following renormalization group 
equations found in Ref. \cite{74Kos}
\begin{eqnarray} &&
\frac{{\rm d}\beta}{{\rm d}\ln(r/a)}=-cy^2 \,, \qquad
\frac{{\rm d}y}{{\rm d}\ln(r/a)}=(2-\beta)y \,,
\label{rg} \end{eqnarray}
where $c$ is a numerical factor of order unity. The $r$-dependent function $y$
is the renormalized fugacity. It determines a concentration of defects
belonging to the bound pairs with separations of the order of $r$, the
concentration can be estimated as $y/r^2$. The renormalized value of $\beta$
determines the dependence of the strength of the Coulomb interaction on the
separation between the charges. In the low-temperature phase, the effective
value of $\beta$ tends to a constant on large scales. The asymptotic value of
$\beta$ is larger than $2$, the critical value $\beta=2$ corresponds to the
transition temperature. In the asymptotic region, where $\beta$ can be
treated as $r$-independent, the renormalized fugacity $y$ remains
$r$-dependent. Its asymptotic behavior can easily be extracted from Eq.
(\ref{rg}):
\begin{eqnarray} &&
y\propto r^{2-\beta} \,.
\label{fug} \end{eqnarray}
Thus, in the low-temperature phase $y$ tends to zero as scale increases.

Let us turn to simultaneous correlation functions of the charge density
$\rho$ (\ref{den}). The odd correlation function are zero. Indeed,
the system is symmetric under permuting defects and antidefects whereas the
charge density (\ref{den}) changes its sign at the permutation. The pair
correlation function can be written as (see, e.g., Ref. \cite{JKKN})
\begin{eqnarray} &&
\langle\rho({\bbox r})\rho(0)\rangle
\sim y^2(r)/r^4 \,.
\label{pair} \end{eqnarray}
A generalization of the relation (\ref{pair}) can be obtained (see Ref.
\cite{Kad}) which is
\begin{eqnarray} &&
\langle\rho({\bbox r}_1)\dots\rho({\bbox r}_{2n})\rangle
\sim  \frac{y^{2n}(r_*)}{r_*^{4n}}\sim
\langle\rho({\bbox r_*})\rho(0)\rangle^n \,,
\label{simul1} \end{eqnarray}
where all separation $|{\bbox r}_i-{\bbox r}_j|$ are assumed to be of the
same order $r_*$. In the large-scale limit where $\beta$ is saturated we have
\begin{eqnarray} &&
\langle\rho(X{\bbox r}_1)\dots\rho(X{\bbox r}_{2n})\rangle
=X^{-2\beta n}\langle\rho({\bbox r}_1)\dots\rho({\bbox r}_{2n})\rangle \,,
\label{high} \end{eqnarray}
where $X$ is an arbitrary factor. The relation (\ref{high}) shows that the
simultaneous statistics of $\rho$ has normal scaling, that is scaling exponents
of the correlation functions of the order $2n$ are equal to $n$ times the
scaling exponent of the pair correlation function (\ref{pair}). We will
demonstrate that the behavior of non-simultaneous correlation functions of the
charge density is quite different.

\subsection{Dynamics}

To examine dynamical characteristics of the system we should formulate a
dynamical equation for a defect motion. Following Ref. \cite{Dynamics1}
we accept the following stochastic equation
\begin{equation}
\frac{{\rm d} {\bbox x}_j}{{\rm d}t}=
-\frac{D}{T}\frac{\partial {\cal F}}
{\partial {\bbox x}_j}+{\bbox\xi}_j \,,
\label{H51} \end{equation}
determining the trajectory of the $j$-th defect. Here ${\cal F}$ is the
free energy (\ref{H04}), $D$ is a diffusion coefficient, and ${\bbox\xi}_j$
are Langevin forces with the correlation function
\begin{equation}
\langle\xi_{i,\alpha}(t_1)\xi_{j,\beta}(t_2)\rangle=
2D\delta_{ij}\delta_{\alpha\beta}\delta(t_1-t_2) \,.
\label{H52} \end{equation}
The diffusion coefficient $D$ determines mobility of the defects. We believe
that the equation (\ref{H51}) is applicable to the dynamics of disclinations
in hexatic films like membranes, freely suspended films and Langmuir films.
The equation for the vortices in superfluid films is a bit more complicated.
It is written in Section \ref{helium} where the correlation functions of the
vorticity are analyzed.

The equations (\ref{H51},\ref{H52}) describe trajectories of separate defects.
We should also take into account annihilation and creation processes. Remember
that we neglect defects with $|n_j|>1$. Next, processes where a number of
defect-antidefect pairs are created at the same point are suppressed since
probability of such events is small due to the energy associated with the cores
of defects. Then we have to take into account the creation processes of single
pairs solely, they are characterized by the creation rate $\bar R(r)$ which is
a probability density for a defect-antidefect pair with the separation $r$
to be created per unit time per unit area. The annihilation processes have to
be characterized by the annihilation rate $R(r)$ which is a probability for a
defect-antidefect pair to annihilate per unit time if the pair is separated
by the distance $r$. Really, both $\bar R(r)$ and $R(r)$ are nonzero only if
$r$ is of the order of the core size $a$. Let us introduce the integrals
\begin{equation}
\bar\lambda=\int{\rm d}^2 r\, \bar R(r) \,, \qquad
\lambda=\int{\rm d}^2 r\, R(r) \,.
\label{rate} \end{equation}
Here, the creation constant $\bar\lambda$ is a probability for a
defect-antidefect pair to be created per unit time per unit area and
$\lambda$ is a constant having the same dimensionality as the diffusion
coefficient $D$. Below, the diffusion coefficient $D$ is
put to unity by rescaling time. Then the annihilation constant $\lambda$ is
a dimensionless parameter of order unity and the creation constant
$\bar\lambda$ can be estimated as
\begin{equation}
\bar\lambda\sim a^{-4} \exp(-2\mu/T) \,,
\label{fuga} \end{equation}
which is the second power of the defect concentration.

As in statics, in dynamics the system of defects can be described in terms of
probability distribution functions. In our case, when the total charge of the
system is zero, only even probability densities ${\cal P}_{2N}(t,{\bbox
x}_1,\dots,{\bbox x}_N,{\bbox z}_1,\dots,{\bbox z}_N)$ are non-zero, where
${\bbox x}_j$ and ${\bbox z}_j$ are positions of the defects and of the
antidefects correspondingly. The total probability should be equal to unity
which gives the normalization condition
\begin{eqnarray} &&
\sum\limits_{N=0}^\infty \frac{1}{(N!)^2}
\int{\rm d}^2x_1\dots{\rm d}^2x_N\,
{\rm d}^2z_1\dots{\rm d}^2z_N\,
\nonumber \\ &&
\times {\cal P}_{2N}(t,{\bbox x}_1,\dots,{\bbox x}_N,
{\bbox z}_1,\dots,{\bbox z}_N)=1 \,.
\label{norm} \end{eqnarray}
Starting from the equation (\ref{H51}) and taking into account the creation
and annihilation processes one can derive a system of master equations for
the probability densities

\end{multicols}

\begin{eqnarray} &&
\frac{\partial{\cal P}_{2N}}{\partial t}
=\sum\limits_j \left\{\frac{\partial}{\partial{\bbox x}_j}
\left[\frac{1}{T}\frac{\partial{\cal F}}{\partial{\bbox x}_j}
{\cal P}_{2N}\right]
+\frac{\partial}{\partial{\bbox z}_j}
\left[\frac{1}{T}\frac{\partial{\cal F}}{\partial{\bbox z}_j}
{\cal P}_{2N}\right]\right\}
+\sum\limits_j\left[\frac{\partial^2}{\partial{\bbox x}_j^2}
+\frac{\partial^2}{\partial{\bbox z}_j^2}\right]{\cal P}_{2N}
\nonumber \\ &&
-\sum\limits_{j,k}R({\bbox x}_j-{\bbox z}_k){\cal P}_{2N}
+\int{\rm d}^2 x\,{\rm d}^2 z\,R({\bbox x}-{\bbox z})
{\cal P}_{2N+2}({\bbox x}_1,\dots,{\bbox x}_N,{\bbox x},
{\bbox z}_1,\dots,{\bbox z}_N,{\bbox z})
\label{master} \\ &&
+\sum\limits_{j,k}\bar R({\bbox x}_j-{\bbox z}_k)
{\cal P}_{2N-2}({\bbox x}_1,\dots,{\bbox x}_{j-1},
{\bbox x}_{j+1},\dots,{\bbox x}_N,
{\bbox z}_1,\dots,{\bbox z}_{k-1},{\bbox z}_{k+1},\dots,{\bbox z}_N)
-\bar\lambda A {\cal P}_{2N} \,,
\nonumber \end{eqnarray}

\begin{multicols}{2}

\noindent
where $A$ is the area of the film.

The Gibbs distribution (\ref{parti}) must be a solution of the master
equations (\ref{master}). The condition imposes the following constraint
on the creation and the annihilation rates
\begin{eqnarray} &&
\bar R(r)=\frac{y_0^2}{a^4}
\left(\frac{a}{r}\right)^{2\beta} R(r) \,,
\label{fdt} \end{eqnarray}
where we imply $r>a$. The constraint (\ref{fdt}) can be treated as the
manifestation of the equilibrium state of the thermal bath which in our case
is related to short-scale fluctuations. Thus the constraint (\ref{fdt}) has
the same origin as Eq. (\ref{H52}). At deriving Eq. (\ref{fdt}) we assumed
that the separation $|{\bbox x}-{\bbox z}|$ in the argument of the rate $R$
or $\bar R$ is smaller than separations between ${\bbox x}$ or ${\bbox z}$
and other points. That is accounted for the small characteristic value of the
separation which are of the order of the core radius $a$. Note that for such
$r$ the factor $y_0(r/a)^{2-\beta}$ entering Eq. (\ref{fdt}) can be treated
as the renormalized value $y$ of the fugacity as follows from Eq. (\ref{rg}).

Principally, the master equations (\ref{master}) enable one to find conditional
probability densities, related to different time moments. Consequently,
starting from the equations one can examine non-simultaneous correlations
in the system. However, there are terms in the master equations (associated
with the creation and annihilation processes) mixing the probability
densities ${\cal P}_{2N}$ with different $N$. That makes the master equations
hardly useful, that is a motivation to look for some more suitable technique.
Such a technique was developed by Doi, Ref. \cite{76Doi}, we formulate it in
the subsequent subsection.

\subsection{Quantum Field Formulation}

The Doi technique \cite{76Doi} enables one to treat systems of classical
particles where creation and annihilation processes occur. The main idea
introduced by Doi is that correlation functions of different quantities
characterizing the particles can be written in the form close to the one known
in the quantum field theory. Of course there are some peculiarities related to
the fact that for classical particles one should deal directly with
probabilities whereas in the quantum field theory one starts from the
scattering matrix. Nevertheless the Doi technique enables, say, to formulate a
diagrammatic expansion with the conventional rules. The technique was
originally developed to describe systems of molecules involved into chemical
reactions. But it is definitely applicable also to the system of point
defects.

The Doi technique is formulated in terms of the creation $\hat\psi$ and
annihilation $\psi$ operators which satisfy the same commutation rules as ones
for Bose-particles
\begin{eqnarray} &&
[\psi({\bbox r}_1),\hat\psi({\bbox r}_2)]=
\delta({\bbox r}_1-{\bbox r}_2)\,,
\nonumber \\ &&
[\hat\psi({\bbox r}_1),\hat\psi({\bbox r}_2)]=
[\psi({\bbox r}_1),\psi({\bbox r}_2)]=0 \,.
\label{do01} \end{eqnarray}
For our system of defects we should introduce annihilation and creation
operators $\psi_\pm$ and $\hat\psi_\pm$ where the subscripts $+$ and $-$ label
fields related to the defects and to the antidefects. The state of the system
at a time moment $t$ can be written in terms of a ``quantum'' state
\begin{eqnarray} &&
|t\rangle=\sum\limits_{N=0}^\infty\frac{1}{(N!)^2}
\int{\rm d}^2\,x_1\dots{\rm d}^2\, x_N\,
{\rm d}^2\,z_1\dots{\rm d}^2\, z_N\,{\cal P}_{2N}
\nonumber \\ &&
\times \hat\psi_+({\bbox x}_1)\dots \hat\psi_+({\bbox x}_N)
\hat\psi_-({\bbox z}_1)\dots \hat\psi_-({\bbox z}_N)|0\rangle \,,
\label{do02} \end{eqnarray}
where ${\cal P}_{2N}$ are the probability densities introduced above and
$|0\rangle$ designates the vacuum state: $\psi_\pm|0\rangle=0$. In accordance
with the expression (\ref{do02}) an evolution of the quantum state $|t\rangle$
is determined by the master equations. The evolution equation can be written as
\begin{eqnarray} &&
\partial_t|t\rangle=-{\cal H}|t\rangle \,, \quad {\rm and} \quad
|t_2\rangle=\exp\left[(t_1-t_2){\cal H}\right]|t_1\rangle \,,
\label{do03} \end{eqnarray}
where ${\cal H}$ is an operator expressed in terms of the fields $\psi_\pm$ and
$\hat\psi_\pm$. By analogy with the quantum field formulation it can be called
the Hamiltonian operator or simply the Hamiltonian.

The total probability must be equal to unity, which leads to Eq. (\ref{norm}).
The condition can be written in terms of the quantum state $|t\rangle$ as
\begin{eqnarray} &&
\langle{\rm sum}|t\rangle=1 \,,
\label{sum} \end{eqnarray}
where
\begin{eqnarray} &&
\langle{\rm sum}|=\langle 0|
\exp\left[\int{\rm d}^2r\,
\left(\psi_++\psi_-\right)\right] \,.
\nonumber \end{eqnarray}
Note the following identity
\begin{eqnarray} &&
\langle{\rm sum}|\hat\psi_\pm({\bbox r})=\langle{\rm sum}|\,,
\label{prop} \end{eqnarray}
which can be easily checked using the commutation rules (\ref{do01}) and the
equality $\langle 0|\hat\psi_\pm=0$. Since the evolution must conserve the
total probability $\langle{\rm sum}|t\rangle$ the relations
\begin{eqnarray} &&
\langle{\rm sum}|{\cal H}=0\,, \quad {\rm and} \quad
\langle{\rm sum}|\exp\left(\tau{\cal H}\right)
=\langle{\rm sum}| \,,
\label{do04} \end{eqnarray}
have to be satisfied, where $\tau$ is an arbitrary parameter.

Quantities characterizing the system can be represented by corresponding
operators, see Ref. \cite{76Doi}. Say, the operator of the charge density is
\begin{eqnarray} &&
\tilde\rho=\hat\psi_+\psi_+-\hat\psi_-\psi_- \,.
\label{do3} \end{eqnarray}
If $\tilde A$ is such operator corresponding to a quantity $A$, then an average
value of the quantity at a time moment $t$ can be expressed as
\begin{eqnarray} &&
\langle A(t) \rangle=
\langle{\rm sum}|\tilde A |t\rangle \,.
\label{do05} \end{eqnarray}
Note that
\begin{eqnarray} &&
\langle\hat\psi_\pm\rangle=\langle{\rm sum}|
\hat\psi_\pm|t\rangle=1 \,,
\label{prop1} \end{eqnarray}
which is a consequence of Eqs. (\ref{sum},\ref{prop}).
The relation (\ref{prop1}) shows that it is natural to shift the creation
operators $\hat\psi_\pm$ introducing new variables (see Ref. \cite{80GS})
\begin{eqnarray} &&
\hat\psi_\pm= 1+\bar\psi_\pm \,, \qquad \langle\bar\psi_\pm\rangle=0 \,.
\label{shift} \end{eqnarray}
Correlation functions of different quantities can be presented analogously to
Eq. (\ref{do05}). For example, the pair correlation function of two
quantities $A(t_1)$ and $B(t_2)$ (we assume $t_2>t_1$) can be written as
\begin{eqnarray} &&
\langle B(t_2)A(t_1)\rangle=
\langle{\rm sum}|\tilde B\exp\left[({t_1}-{t_2})
{\cal H}\right]\tilde A|t_1\rangle \,.
\label{do1} \end{eqnarray}
Using the relations (\ref{do03},\ref{do04}) we can rewrite the expression
(\ref{do1}) as
\begin{eqnarray} &&
\langle B(t_2)A(t_1)\rangle=
\langle{\rm sum}|\tilde B(t_2)\tilde A(t_1) |{\rm in}\rangle \,,
\label{do06} \end{eqnarray}
where $\tilde A(t),\tilde B(t)$ in Eq. (\ref{do06}) are operators in
the Heisenberg representation
\begin{eqnarray} &&
\tilde A(t)=\exp\left[-({t_{\rm f}}-{t}){\cal H}\right]
\tilde A \exp\left[-({t}-t_{\rm in}){\cal H}\right] \,,
\label{heis} \end{eqnarray}
satisfying the equation
$\partial_t \tilde A(t)=\left[{\cal H},\tilde A(t)\right]$. In Eq.
(\ref{do06}) $|{\rm in}\rangle$ is an initial state (realized at a time
moment $t_{\rm in}$) and $t_{\rm f}$ is a ``final'' time, so that
$t_{\rm f}>t_2>t_1>t_{\rm in}$.

The expressions like (\ref{do06}) enable one to reformulate the problem of
calculating correlation functions in terms of a functional integral, see Ref.
\cite{85Pel}. Namely, we can write
\begin{eqnarray} &&
\langle A_1(t_1)\dots A_n(t_n)\rangle
=\int {\cal D}\hat\psi_\pm {\cal D}\psi_\pm
\tilde A_1\dots \tilde A_n
\nonumber \\ &&
\times\exp\biggl\{-\int_{-\infty}^{t_{\rm f}}{\rm d}t\, \left[{\cal H}
+\int {\rm d}^2 r\,\left(\hat\psi_+\partial_t\psi_+
+\hat\psi_-\partial_t\psi_-\right)\right]
\nonumber \\ &&
+\int {\rm d}^2 r\,\left[\psi_+(t_{\rm f},{\bbox r})
+\psi_-(t_{\rm f},{\bbox r})\right] \biggr\} \,,
\label{path1} \end{eqnarray}
where $\psi_\pm, \hat\psi_\pm$ are to be interpreted as functions of $t$ and
${\bbox r}$. We assume that $t_{\rm f}>t_1,\dots,t_n$ in Eq. (\ref{path1}).
Deriving the expression one has taken the limit $t_{\rm in}\to-\infty$ and
assumed $|{\rm in}\rangle=|0\rangle$. Because of the creation processes the
vacuum has to be turned into a stationary state during the infinite time. To
ensure convergence of the functional integral (\ref{path1}) the integration
contour over the field $\hat\psi$ should go parallel to the imaginary axis.
Note that the shift (\ref{shift}) kills the boundary term
$\int{\rm d}^2r\,(\psi_++\psi_-)$: It is cancelled by a contribution
originating from the derivatives $\partial_t\psi_\pm$ after integrating over
time. Then we come to a conventional representation of the correlation
functions in terms of a functional integral
\begin{eqnarray} &&
\langle A_1(t_1)\dots A_n(t_n)\rangle
=\int {\cal D}\bar\psi_\pm {\cal D}\psi_\pm
\exp\biggl\{-\int{\rm d}t\, \biggl[{\cal H}
\nonumber \\ &&
+\int {\rm d}^2 r\,\left(\bar\psi_+\partial_t\psi_+
+\bar\psi_-\partial_t\psi_-\right)\biggr]\biggr\}
\tilde A_1\dots \tilde A_n \,.
\label{path} \end{eqnarray}
The relation (\ref{path1}) or (\ref{path}) is a convenient starting point for
treating a system of particles involved into the creation and annihilation
processes.

\section{Diagrammatic Representation}
\label{quantum}

Below, we apply the Doi technique to our particular problem. The explicit
expression for the Hamiltonian determining the evolution of the defect system
in accordance with Eq. (\ref{do03}) can be found from the master equations
(\ref{master}). Comparing the equations with Eqs. (\ref{do02},\ref{do03})
we get
\begin{eqnarray} &&
{\cal H}={\cal H}_0+{\cal H}_R+{\cal H}_\beta \,.
\label{ham} \end{eqnarray}
The explicit expressions for the terms entering Eq. (\ref{ham}) are
\begin{eqnarray} &&
{\cal H}_0=\int{\rm d}^2 r\,\left(
\nabla\hat\psi_+\nabla\psi_+
+\nabla\hat\psi_-\nabla\psi_-\right)
\label{ham0}  \end{eqnarray}
\begin{eqnarray} &&
{\cal H}_R=-\int{\rm d}^2 r_1\,{\rm d}^2 r_2\,
\biggl[\bar R({\bbox r}_1-{\bbox r}_2)
(\hat\psi_{+,1}\hat\psi_{-,2}-1)
\nonumber \\ &&
+R({\bbox r}_1-{\bbox r}_2)
(\psi_{+,1}\psi_{-,2}
-\hat\psi_{+,1}\hat\psi_{-,2}\psi_{+,1}\psi_{-,2})\biggr]
\label{hamr} \end{eqnarray}
\begin{eqnarray} &&
{\cal H}_\beta=2\beta\int {\rm d}^2 r_1\,{\rm d}^2 r_2\,
\left(\nabla\hat\psi_{+,1}\hat\psi_{-,2}
-\hat\psi_{+,1}\nabla\hat\psi_{-,2}\right)
\nonumber \\ &&
\times \frac{{\bbox r}_1-{\bbox r}_2}{|{\bbox r}_1-{\bbox r}_2|^2}
\psi_{+,1}\psi_{-,2}
\nonumber \\ &&
-2\beta\int {\rm d}^2 r_1\,{\rm d}^2 r_2\,\biggl[
\nabla\hat\psi_{+,1}\hat\psi_{+,2}
\frac{{\bbox r}_1-{\bbox r}_2}{|{\bbox r}_1-{\bbox r}_2|^2}
\psi_{+,1}\psi_{+,2}
\nonumber \\ &&
+\nabla\hat\psi_{-,1}\hat\psi_{-,2}
\frac{{\bbox r}_1-{\bbox r}_2}{|{\bbox r}_1-{\bbox r}_2|^2}
\psi_{-,1}\psi_{-,2}\biggr] \,,
\label{hamb} \end{eqnarray}
where $\psi_{+,1}=\psi_+(t,{\bbox r}_1)$ and so further. The diffusive
contribution (\ref{ham0}) is related to the Langevin forces, in Eq.
(\ref{hamr}) $R$ is the annihilation rate and $\bar R$ is the creation rate
for the defect-antidefect pairs (the quantities were introduced in Section
\ref{basic}), and the term (\ref{hamb}) describe the Coulomb interaction.
Using the property (\ref{prop}), one can easily check the conditions
(\ref{do04}) for all contributions (\ref{ham0}-\ref{hamb}).

Performing the substitution (\ref{shift}) we can express the Hamiltonian
(\ref{ham}) in terms of the fields $\bar\psi_\pm$. Note that the shift
(\ref{shift}) kills terms of the second order proportional to $\lambda$,
$\bar\lambda$ and generates additional third-order vertices. Of course one
can work in both representations. It is more convenient for us to use Eqs.
(\ref{path1},\ref{ham}). We can easily convince ourselves that odd
correlation functions of the charge density (\ref{do3}) are zero. Indeed, 
the exponent in Eq. (\ref{path1}) is invariant under permuting
$\psi_+\leftrightarrow\psi_-$, $\hat\psi_+\leftrightarrow\hat\psi_-$, whereas
the charge density changes its sign at the permutation. The constraint
(\ref{Ch1}) shows that the symmetry is not spontaneously broken what could
lead to non-zero odd correlation functions.

It follows from Eq. (\ref{heis}) that the commutator $[{\cal H},\tilde\rho]$
should be equal to $-\nabla{\bbox j}$ where ${\bbox j}$ is the current density
operator. Calculating the commutator with Eqs. (\ref{do3},\ref{ham}) we get
\begin{eqnarray} &&
{\bbox j}=\nabla\hat\psi_+\psi_+-\hat\psi_+\nabla\psi_+
-\nabla\hat\psi_-\psi_-+\hat\psi_-\nabla\psi_-
\nonumber \\ &&
-(\hat\psi_+\psi_++\hat\psi_-\psi_-)\nabla\phi \,,
\label{curr} \end{eqnarray}
where $\phi$ is an ``electrostatic potential''
\begin{eqnarray} &&
\phi({\bbox r})=-2\beta\int{\rm d}^2 x\,
\ln\left(\frac{|{\bbox r}-{\bbox x}|}{a}\right)
\tilde\rho({\bbox x}) \,.
\label{poten} \end{eqnarray}
Principally, besides the ``internal'' potential (\ref{poten}) an ``external''
potential $\phi_{\rm ext}$ can be imposed onto the system, satisfying the
equation $\nabla^2\phi_{\rm ext}=0$. Then an ``external force'' should be
added to the right-hand side of the equation (\ref{H51}). The force generates
an ``external'' contribution to the Hamiltonian
\begin{eqnarray} &&
{\cal H}_{\rm ext}=\int{\rm d}^2 r\,\left(
\nabla\hat\psi_+\psi_+
-\nabla\hat\psi_-\psi_-\right)\nabla\phi_{\rm ext} \,.
\label{ext} \end{eqnarray}
The expression (\ref{curr}) for the current density has also to be corrected
by substituting $\phi\to\phi+\phi_{\rm ext}$. With the term (\ref{ext}) we
can examine susceptibilities describing a response of the system to the
external influence.

Substituting the expression (\ref{ham}) into (\ref{path1}) or (\ref{path})
and expanding the exponent over ${\cal H}_R$ and ${\cal H}_\beta$ one can
obtain a conventional perturbation series for calculating different
correlation functions of $\psi,\hat\psi$. The series is an expansion over
$R$, $\bar R$ and $\beta$ in terms of the conventional diffusion propagators:
\begin{eqnarray} &&
G(t,{\bbox r})=\langle \psi_+(t,{\bbox r})\hat\psi_+(0,0)\rangle_0
\nonumber \\ &&
= \langle \psi_-(t,{\bbox r})\hat\psi_-(0,0)\rangle_0
=\frac{\theta(t)}{4\pi t}\exp\left(-\frac{r^2}{4t}\right) \,,
\label{diff} \end{eqnarray}
where $\theta(t)$ is the step function. However, effects related to the
Coulomb interaction and to the annihilation processes are not weak. Therefore
one must take into account the Coulomb interaction and the annihilation
processes exactly. By other words, at calculating the correlation functions
one must consider the complete series over $\beta$ and $R$. Fortunately, the
the expansion over $\bar R$ is equivalent to an expansion over the fugacity
$y$ which is assumed to be a small parameter. Therefore we can take only
principal terms in the expansion over $\bar R$.

\begin{figure}
\centerline{\psfig{figure=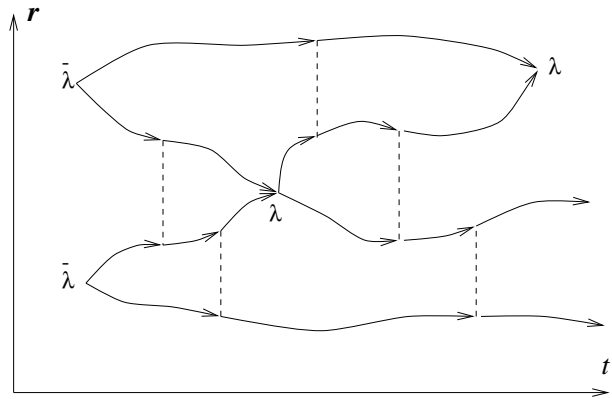,width=8cm}}
\caption{Typical diagram block.}
\label{diag}
\end{figure}

The perturbation expansion can be formulated as a diagrammatic series. We
develop the diagrammatic technique starting from the representation
(\ref{path1}), pushing the final time $t_{\rm f}$ to the far future. We
depict the propagator (\ref{diff}) by a line directed from $\hat\psi$ to
$\psi$. The term with the creation rate $\bar R$ in Eq. (\ref{ham}) generates
vertices where two propagator lines start, the vertices correspond to the
defect-antidefect creation processes. The Coulomb term in Eq.  (\ref{ham})
generates two-point objects which we will designate by dashed lines, such
line describes the Coulomb interaction of defects located in points connected
by the line. And the term proportional to the annihilation rate $R$ in Eq.
(\ref{ham}) produces two types of vertices. First, it produces vortices where
two propagator lines finish, that corresponds to an annihilation process.
Second, it produces fourth-order vertices which correspond to an effective
interaction related to a finite probability for a defect-antidefect pair to
annihilate, see Ref. \cite{76Doi}. A typical diagram block is presented in
Fig. \ref{diag}. The block is drawn in real ${\bbox r}-t$ space-time. The
curves constituted of the propagator lines can be interpreted as trajectories
of defects and antidefects. Due to causality the particles always move
forward in time. Note that the dashed lines corresponding to the Coulomb
interaction are perpendicular to the $t$-axis since the interaction is
simultaneous.

\subsection{Pair Conditional Probability}

In the subsection we examine an auxiliary object which will be needed for us
at intermediate stages of subsequent calculations. The object is the
following correlation function
\begin{eqnarray} &&
M(t_2-t_1,{\bbox r}_1,{\bbox r}_2,{\bbox r}_3,{\bbox r}_4)
\nonumber \\ &&
=\langle \psi_+(t_2,{\bbox r}_1)\psi_-(t_2,{\bbox r}_2)
\hat\psi_+(t_1,{\bbox r}_3)\hat\psi_-(t_1,{\bbox r}_4)\rangle \,.
\label{mmm} \end{eqnarray}
For a stationary case the average (\ref{mmm}) depends on the difference
$t=t_2-t_1$ only. Due to causality $M$ is equal to zero provided $t<0$. The
quantity (\ref{mmm}) can be interpreted as a probability density to find a
defect and an antidefect at the time moment $t_2$ in the points ${\bbox r}_1$
and ${\bbox r}_2$ provided they were located in the points ${\bbox r}_3$ and
${\bbox r}_4$ at the time moment $t_1$. It can be considered also as a
two-particle matrix element of the evolution operator
$\exp[-(t_2-t_1){\cal H}]$.

\begin{figure}
\centerline{\psfig{figure=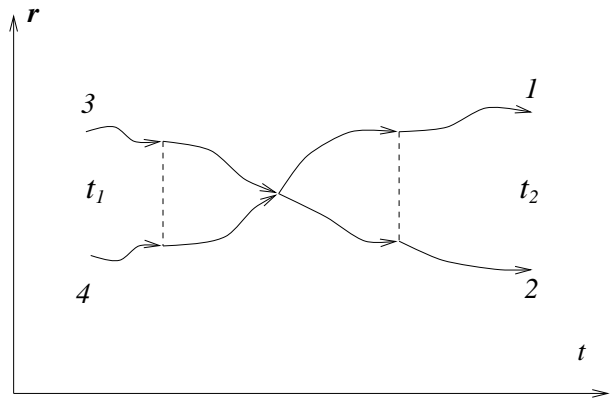,width=8cm}}
\caption{Typical diagram for the conditional probability $M$.}
\label{diag1}
\end{figure}

As we explained above, the perturbation series in terms of the creation rate
$\bar R$ is an expansion over a small parameter. Here we examine the
principal contribution to the conditional probability (\ref{mmm}) which is of
the zero order over $\bar R$. Then the average (\ref{mmm}) can be represented
as a series of diagrams of the type depicted in Fig. \ref{diag1}. One can
interpret the picture as trajectories of a defect and of an antidefect which
are driven by the Langevin forces, and are influenced the Coulomb interaction
(dashed lines) and the effective interaction associated with the annihilation
processes (point vertex). Note that in this approximation direct annihilation
events do not contribute to the conditional probability (\ref{mmm}) since
they would lead to terminating the lines in the diagrams.

It is of crucial importance that both the Coulomb interaction and the effective
interaction associated with the annihilation processes are local in time.
Therefore all the diagrams representing the conditional probability (\ref{mmm})
are ladder diagrams, like in Fig. \ref{diag1}. Summing up the ladder sequence
we get an equation for $M$ which can be written in the differential form
\begin{eqnarray} &&
\partial_t M=\left(\nabla_1^2+\nabla_2^2\right)M
+2\beta\left(\nabla_1-\nabla_2\right)\left[
\frac{{\bbox r}_1-{\bbox r}_2}{|{\bbox r}_1-{\bbox r}_2|^2}M\right]
\nonumber \\ &&
-R({\bbox r}_1-{\bbox r}_2)M
+\delta(t)\delta\left({\bbox r}_1-{\bbox r}_3\right)
\delta\left({\bbox r}_2-{\bbox r}_4\right) \,.
\label{MM} \end{eqnarray}
Since $M=0$ at $t<0$ we conclude from Eq. (\ref{MM}) that at $t\to+0$
\begin{equation}
M(t,{\bbox r}_1,{\bbox r}_2,{\bbox r}_3,{\bbox r}_4)
\to \delta\left({\bbox r}_1-{\bbox r}_3\right)
\delta\left({\bbox r}_2-{\bbox r}_4\right) \,.
\label{small} \end{equation}
The total probability to find the defect-antidefect pair is determined by the
integral $\int {\rm d}^2 r_1\,{\rm d}^2 r_2\,M$. Let us calculate the time
derivative of this integral substituting $\partial_t M$ by the right-hand
side of Eq. (\ref{MM}). Then the first two terms will give zero contributions
(since they are total derivatives) and only the term with $R$ will produce a
non-zero (negative) contribution. It is quite natural since the Langevin
forces and the Coulomb interaction cannot change the total probability
whereas the annihilation processes diminish it. Note that all the terms in
the right-hand side of Eq. (\ref{MM}) proportional to $M$ have the same
dimensionality. Therefore one could expect a simple scaling behavior when
$t$ scales as $r^2$. The subsequent calculations confirm the expectation.

In terms of the variables
\begin{eqnarray} &&
{\bbox r}={\bbox r}_1-{\bbox r}_2 \,, \qquad
{\bbox\varrho}=\frac{{\bbox r}_1+{\bbox r}_2}{2}\,, \qquad
{\bbox r}_0={\bbox r}_3-{\bbox r}_4 \,,
\label{J3} \end{eqnarray}
the equation (\ref{MM}) for $M$ is rewritten as
\begin{eqnarray} &&
\partial_t M=\left(\frac{1}{2}\nabla_{\varrho}^2
+2\nabla_r^2+4\beta\nabla_r\frac{\bbox r}{r^2}-R({\bbox r})\right)M
\nonumber \\ &&
+\delta(t)\delta\left({\bbox r}-{\bbox r}_0\right)
\delta\left({\bbox\varrho}-{\bbox r}_3/2-{\bbox r}_4/2\right) \,.
\nonumber \end{eqnarray}
We see that the differential operator in the right-hand side of the equation
falls into two parts depending on ${\bbox\varrho}$ and ${\bbox r}$ only and
that the ``source'' is a product of $\delta$-functions of the same variables.
Therefore the solution of the equation can be written in a multiplicative form
\begin{eqnarray} &&
M=\frac{1}{2\pi t}
\exp\left\{-\frac{\left(2{\bbox\varrho}-{\bbox r}_3
-{\bbox r}_4\right)^2}{8t}\right\}
S(t,{\bbox r},{\bbox r}_0) \,,
\label{H65} \end{eqnarray}
the function $S$ satisfies the following equation
\begin{eqnarray} &&
\partial_t S=2\nabla^2S+4\beta\nabla\left(\frac{\bbox r}{r^2}S\right)
\nonumber \\ &&
-R({r})S+\delta(t)\delta\left({\bbox r}-{\bbox r}_0\right) \,.
\label{doo} \end{eqnarray}
Note that
\begin{eqnarray} &&
{\rm if} \quad t\to+0 \quad {\rm then}
\quad S\to \delta\left({\bbox r}-{\bbox r}_0\right) \,.
\label{small1} \end{eqnarray}
The relation follows from Eq. (\ref{doo}) and causality (leading to $S=0$ for
negative $t$).

In accordance with Eq. (\ref{H65}), a motion of the mass center and the
relative motion of the defects are separated. The motion of the mass center is
purely diffusive whereas the relative motion is strongly influenced by the
interaction. The function $S$ can be treated as the probability density for the
relative motion of the defect-antidefect pair. It is natural to expand the
function into the Fourier series over the angle $\varphi$ between the vectors
${\bbox r}$ and ${\bbox r}_0$:
\begin{eqnarray} &&
S(t,{\bbox r},{\bbox r}_0)
=\sum\limits_{-\infty}^{+\infty}S_m(t,r,r_0)\exp(im\varphi) \,.
\label{angle} \end{eqnarray}
Motions corresponding to different angular harmonics are separated.
In terms of the angular harmonics Eq. (\ref{doo}) is rewritten as
\begin{eqnarray} &&
\frac{1}{2}\partial_t S_m
=\left[\partial_r^2+(1+2\beta)\frac{1}{r}\partial_r
-\frac{m^2}{r^2}\right]S_m
\nonumber \\ &&
-\frac{1}{2}R(r)S_m
+\frac{1}{4\pi r_0}\delta(t)\delta(r-r_0) \,.
\label{four} \end{eqnarray}
It is possible to get equations for $S$ analogous to Eqs.
(\ref{doo},\ref{four}) in terms of ${\bbox r}_0$. They have practically the
same form as Eqs. (\ref{doo},\ref{four}). The only difference is in the sign
of $\beta$ which is opposite. That leads to the relation
\begin{eqnarray} &&
S_m(t,r,r_0)=\left(\frac{r_0}{r}\right)^{2\beta}
S_m(t,r_0,r) \,.
\label{rr0} \end{eqnarray}
Let us stress that the relation (\ref{rr0}) is correct for an arbitrary
function $R(r)$.

Consider a behavior of the angular harmonics $S_m(t,r,r_0)$ at small $r$.
More precisely, we assume $t\gg a^2$ and examine the region
$\sqrt t\gg r\gg a$. Then it is possible to use the equation (\ref{four})
with the time derivative and the annihilation term neglected. As a result we
get
\begin{eqnarray} &&
S_m=C_{1,m}r^{\nu-\beta}
+C_{2,m}r^{\nu-\beta}(r/a)^{-2\nu} \,,
\label{att} \\ &&
\nu=\sqrt{\beta^2+m^2} \,,
\label{att1} \end{eqnarray}
where $C_{1,m},C_{2,m}$ are some factors dependent on $t$ and $r_0$. The
ratio of the factors is determined by a concrete $r$-dependence of the
annihilation rate $R$, one can assert only that $C_{1m}$ and $C_{2m}$ are of
the same order. Therefore, if we consider the behavior of the function $S$
for $r\gg a$, then the second term in the right-hand side of Eq. (\ref{att})
can be neglected. By other words, being interested in the scales $r\gg{a}$,
we can solve the equation (\ref{four}) neglecting the annihilation term and
requiring a finite value of $S_m$ at $r\to 0$ instead.

The requirement can be treated as the boundary condition for $S_m$ at small
$r$. The other boundary condition is that $S_m$ tend to zero at $r\to\infty$.
The equations for $S_m$ with the boundary conditions are solved in Appendix
\ref{condi}. We present here only the answer
\begin{equation}
S_m(t,r,r_0)=\frac{1}{8\pi t}\left(\frac{r_0}{r}\right)^\beta
\exp\left(-\frac{r^2+r_0^2}{8t}\right)
I_\nu\left(\frac{rr_0}{4t}\right) \,,
\label{H69} \end{equation}
where $I$ is the modified Bessel function and $\nu$ is introduced by Eq.
(\ref{att1}). Remind that the expression is correct provided
$r,r_0,\sqrt t\gg a$. Extracting from Eq. (\ref{H69}) an asymptotics at small
$r$ we get
\begin{eqnarray} &&
C_{1,m}=\frac{r_0^\beta}{8\pi\Gamma(1+\nu)t}
\left(\frac{r_0}{8t}\right)^\nu
\exp\left(-\frac{r_0^2}{8t}\right) \,.
\nonumber \end{eqnarray}

Note that the Coulomb term in Eq. (\ref{doo}) produces a probability flux to
the origin. To find it we should integrate the equation (\ref{doo}) over a disk
of a radius $a\ll r\ll\sqrt t$ centered at the origin and single out the
contribution to $\partial_t\int{\rm d}^2r S$ associated with the Coulomn term.
Then we find the flux $\lambda_{\rm r}C_{1,0}$ where
\begin{eqnarray} &&
\lambda_{\rm r}=8\pi\beta \,.
\label{rrr} \end{eqnarray}
One can treat the quantity (\ref{rrr}) as the renormalized (``dressed'') value
of the annihilation constant. Now we understand why the solution (\ref{H69})
(realized at $r\gg{a}$) is insensitive to a particular form of the annihilation
rate. The probability for a defect-antidefect pair with the separation $r\gg a$
to annihilate is determined by the Coulomb attraction. And only the behavior
of the probability density at $r\sim{a}$ is sensitive to the particular form
of the annihilation rate $R(r)$: The coefficients $C_{2,m}$ in Eq.
(\ref{att}) are positive if $\lambda<\lambda_{\rm r}$ and are negative if
$\lambda>\lambda_{\rm r}$.

Returning to the conditional probability (\ref{mmm}) we obtain
from Eqs. (\ref{H65},\ref{angle},\ref{H69})
\begin{eqnarray} &&
M=\frac{1}{(4\pi t)^2}\left(\frac{r_0}{r}\right)^\beta
\exp\left\{-\frac{\left({\bbox r}_1+{\bbox r}_2-{\bbox r}_3
-{\bbox r}_4\right)^2}{8t}\right\}
\nonumber \\ &&
\times\exp\left(-\frac{r^2+r_0^2}{8t}\right)
\sum\limits_{m=-\infty}^{+\infty}\exp(im\varphi)\
I_\nu\left(\frac{rr_0}{4t}\right)\,,
\label{H70} \end{eqnarray}
where $\nu$ is introduced by Eq. (\ref{att1}). One can easily check that the
expression (\ref{H70}) is reduced to Eq. (\ref{small}) at $t\to+0$.  It is
possible to calculate an explicit expression for the total probability to
find a defect and an antidefect in any points at a fixed time separation $t$,
see Appendix \ref{condi}. The asymptotic behavior of the expression
(\ref{H70}) at the condition $rr_0/t\gg1$ is examined in Appendix
\ref{asympt}, see Eq. (\ref{WW8}). If both $r,r_0$ are much greater than
$\sqrt t$, it can be written as
\begin{eqnarray} &&
M\approx\frac{1}{(4\pi t)^2}
\exp\left\{-\frac{({\bbox r}-{\bbox r}_0)^2}{8t}\right\}
\nonumber \\ &&
\times \exp\left\{-\frac{\left({\bbox r}_1+{\bbox r}_2-{\bbox r}_3
-{\bbox r}_4\right)^2}{8t}\right\} \,.
\label{W8} \end{eqnarray}
The answer is quite natural. The characteristic values of the separations
${\bbox r}_1-{\bbox r}_3$ and of ${\bbox r}_2-{\bbox r}_4$ are of the order of
$\sqrt t$ and are consequently much smaller than $|{\bbox r}_1-{\bbox r}_2|$
(or $|{\bbox r}_3-{\bbox r}_4|$). Then $r\approx r_0$ and it is possible to
neglect all terms, containing $|{\bbox r}_1-{\bbox r}_2|$ in the denominators,
in the equation (\ref{MM}). Thus we come to a purely diffusive equation leading
to the asymptotic law (\ref{W8}).

The consideration presented above can be generalized for the case when an
``external electrostatic potential'' $\phi_{\rm ext}$ is imposed onto the
system. Its influence is described by the contribution (\ref{ext}) to the
Hamiltonian. Performing the same procedure as above we get a modified
equation for the correlation function (\ref{mmm})
\begin{eqnarray} &&
\frac{\partial}{\partial t_2}M
=\left(\nabla_1^2+\nabla_2^2\right)M
+4\beta\nabla_r
\left(\frac{{\bbox r}}{r^2}M\right)-R(r)M
\nonumber \\ &&
+\nabla\phi_{\rm ext}(t_2,{\bbox r}_1)\nabla_1M
-\nabla\phi_{\rm ext}(t_2,{\bbox r}_2)\nabla_2M
\nonumber \\ &&
+\delta(t_2-t_1)\delta\left({\bbox r}_1-{\bbox r}_3\right)
\delta\left({\bbox r}_2-{\bbox r}_4\right) \,.
\label{MM1} \end{eqnarray}
The expression (\ref{MM1}) shows that the gradient of the external potential
has to be added to the gradient of the internal one. Of course, in the
presence of the external field the correlation function (\ref{mmm}) depends
on both time moments $t_1$ and $t_2$. Note that the operator in the
right-hand side of Eq. (\ref{MM1}) is the same as that for the Fokker-Plank
equation formulated in Ref. \cite{Dynamics1} (excluding for the annihilation
term).

\section{Renormalization}
\label{renorm}

In this section we are going to discuss effects related to high-order terms
over the creation rate $\bar R$. The effects are relevant only near the
transition point where $\beta$ is close to $2$. Then the influence of
small-scale defect-antidefect pairs on larger scales becomes essential. In the
situation the most natural language is the renormalization group approach. One
can formulate a renormalization group procedure in the spirit of Kosterlitz, 
Ref. \cite{74Kos}. We will single out blocks corresponding to small 
separations of the pairs and treat them as renormalized quantities entering 
the Hamiltonian (\ref{ham}).

\subsection{Creation and annihilation rates}

\begin{figure}
\centerline{\psfig{figure=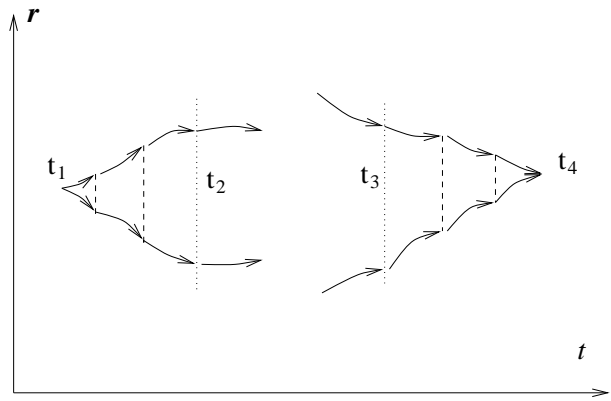,width=8cm}}
\caption{Vicinities of Creation and Annihilation Points}
\label{diag2}
\end{figure}

Sizes of the pairs are small near creation and near annihilation points.
Here, we consider vicinities of the points. Then it is possible to neglect
the interaction of the defect and of the antidefect with the environment.
Thus we turn to the situation when only a single pair can be treated. If this
is the case then one should analyze diagram blocks of the type drawn in Fig.
\ref{diag2}. The left part of the figure corresponds to a vicinity of the
creation occurring at a time moment $t_1$ and the right part of the figure
corresponds to a vicinity of the annihilation occurring at a time moment
$t_4$.

Consider processes occurring during a time interval $\tau$ from the creation
time $t_1$. One can separately treat a block corresponding to the time interval
from $t_1$ till $t_2=t_1+\tau$. For the purpose we use the well-known property
of the propagators (\ref{diff})
\begin{eqnarray} &&
G(s_3-s_1,{\bbox r})=\int{\rm d}^2 x\,
G(s_3-s_2,{\bbox r}-{\bbox x})
\nonumber \\ &&
\times G(s_2-s_1,{\bbox x}) \,,
\label{conv} \end{eqnarray}
where $s_3>s_2>s_1$. For each diagram we extract propagators $G$ containing
$t_2$ inside their time interval and represent the propagators like in Eq.
(\ref{conv}) believing $s_2=t_2$. The procedure is reflected in Fig.
\ref{diag2} where the dotted line represents a plane $t=t_2$ in the
${\bbox r}-t$ space-time and the integration in Eq. (\ref{conv}) corresponds
to the integration in the plane. As a result, the block to the left of the
plane is separated, it is characterized by the time separation $\tau$ and by
two points ${\bbox r}_1$ and ${\bbox r}_2$ lying in the plane, the points are
intersections of the plane with the trajectories of the particles. The block
has to be inserted into more complicated objects via a convolution over
${\bbox r}_1$ and ${\bbox r}_2$.

The same is true for the vicinity of the annihilation point also. Let us take
a time moment $t_3$ separated by a time interval $\tau\gg a^2$ from an
annihilation time $t_4$. Then it is possible to introduce the block which is
a sum of the diagrams where the trajectories of the annihilating particles
start from two given points ${\bbox r}_3$ and ${\bbox r}_4$ at $t=t_3$. The
block has to be inserted into more complicated objects via a convolution over
the points. In the vicinity of the annihilation point we can take into
account the interaction of the annihilating defect-antidefect pair solely.
That leads to the same ladder diagrams treated in Section \ref{quantum}.
Therefore we can write an expression for the block without an additional
analysis
\begin{eqnarray} &&
R_\tau({r}_0)=\int{\rm d}^2 r_1\,{\rm d}^2 r_2\,R({\bbox r})
M(\tau,{\bbox r}_1,{\bbox r}_2,{\bbox r}_3,{\bbox r}_4)
\nonumber \\ &&
=\int{\rm d}^2r\, S(\tau,{\bbox r},{\bbox r}_0)R({r}) \,.
\label{dress} \end{eqnarray}
Here ${\bbox r}$ and ${\bbox r}_0$ are defined by Eq. (\ref{J3}), $M$ is the
conditional probability (\ref{mmm}), $S$ is the conditional probability for the
relative motion of the defects, see Eq. (\ref{H65}), it is the solution of
the equation (\ref{doo}). The physical meaning of the quantity
$R_\tau({r}_0)$ is a distribution of the annihilating particles over the
separation $r_0$ between the particles at the time moment $t_3$. It is
natural to name this distribution ``dressed'' annihilation rate since the
quantity determines a probability for the particles to annihilate after the
time interval $\tau$. Note that all processes occurring on scales larger than
$\sqrt\tau$ are sensitive only to this dressed quantity.

Let us substitute into Eq. (\ref{dress}) the product $RS$ expressed from Eq.
(\ref{doo}). The terms with the total derivatives give zero contribution to
the integral over ${\bbox r}$ and we get
\begin{eqnarray} &&
R_\tau({r}_0)=-\partial_\tau
\int{\rm d}^2r\,S(\tau,{\bbox r},{\bbox r}_0) \,,
\label{dres1} \end{eqnarray}
Since $S(\tau)$ tends to zero at $\tau\to+\infty$ and is zero for negative
$\tau$ we get from (\ref{small1},\ref{dres1})
\begin{eqnarray} &&
\int{\rm d}\tau\,R_\tau({r}_0) =1 \,.
\label{unity} \end{eqnarray}
The relation means that the total probability for a given pair to annihilate
is equal to unity. As is seen from Eq. (\ref{att}) at the condition
$\tau\gg a^2$ the main contribution to the integral in the right-hand side of
Eq. (\ref{dres1}) is associated with the region $r\sim\sqrt\tau$ and
therefore the contribution to the integral associated with the region
$r\sim{a}$ is negligible. Therefore we can use the expression (\ref{angle})
with Eq. (\ref{H69}). Substituting the expression (\ref{dres2}) into Eq.
(\ref{dres1}) we get a universal expression for the dressed quantity
$R_\tau({r}_0)$ which is insensitive to the bare quantity $R(r)$.

In Sec. \ref{quantum} we established the renormalized value (\ref{rrr}) of the
annihilation constant $\lambda$. This analysis concerned the fourth-order
interaction term written in Eq. (\ref{ham}). Below we demonstrate that the
renormalized coefficient at the second-order annihilation term has the same
value, independent of the bare one. In accordance with Eq. (\ref{rate}), to
find the renormalized value $\lambda_r$ we should calculate the integral of
$R_\tau({r}_0)$. As is demonstrated in Appendix \ref{condi} at $\tau\gg a^2$
the value of the integral is independent of $\tau$ and coincides with the
value written in Eq. (\ref{rrr}), as one anticipated:
\begin{eqnarray} &&
\lambda_{\rm r}=\int {\rm d}^2r_0\, R_\tau({r}_0)
=8\pi\beta \,.
\label{dres3} \end{eqnarray}
The phenomenon resembles the renormalization of the reaction rate due to
diffusion, see Refs. \cite{76Doi,86Pel}.

Analogously, one can introduce the renormalized creation rate
$\bar R_\tau(r)$ which is determined by the block describing the vicinity of
the creation point (see Fig. \ref{diag2}). Summing up the same ladder sequence
of the diagrams we get
\begin{eqnarray} &&
\bar R_\tau({\bbox r})=
\int{\rm d}^2 r_3\,{\rm d}^2 r_4\,\bar R({\bbox r}_0)
M(\tau,{\bbox r}_1,{\bbox r}_2,{\bbox r}_3,{\bbox r}_4)
\nonumber \\ &&
=\int{\rm d}^2 r_0\,\bar R({\bbox r}_0)
S(\tau,{\bbox r},{\bbox r}_0) \,.
\label{J1} \end{eqnarray}
Here ${\bbox r}$ and ${\bbox r}_0$ are defined by Eq. (\ref{J3}), $M$ is the
conditional probability (\ref{mmm}), $S$ is the conditional probability for
the relative motion of the defects, see Eq. (\ref{H65}). Using the relations
(\ref{fdt},\ref{rr0}) we get from Eq. (\ref{J1})
\begin{eqnarray} &&
\bar R_\tau({\bbox r})
=\frac{y_0^2}{a^4}\left(\frac{a}{r}\right)^{2\beta}
R_\tau(r) \,.
\label{fdt1} \end{eqnarray}
Thus we see that the relation (\ref{fdt}) is reproduced for the renormalized
quantities $R_\tau$ and $\bar R_\tau$.

The renormalized creation rate $\bar R_\tau({\bbox r})$ can be interpreted as
a probability density to find a defect-antidefect pair with a space
separation ${\bbox r}$ provided the pair was born on time separated by $\tau$
from the measurement. Let us calculate the total probability density
$\bar\lambda_{\rm r}$ to find the defect-antidefect pair at a fixed time
separation $\tau$ regarding $\tau\gg a^2$. The probability is determined by
the integral of $\bar R_\tau({\bbox r})$ over ${\bbox r}$. We conclude from
the expressions (\ref{H69},\ref{dres1}) that the integral is determined by
the region $r\sim\sqrt\tau$. Taking into account Eq. (\ref{dres3}) we get
\begin{eqnarray} &&
\bar\lambda_{\rm r}=\int{\rm d}^2 r\,\bar R_\tau({\bbox r})
\sim\frac{y_0^2}{a^4}
\left(\frac{a^2}{\tau}\right)^{\beta} \,.
\label{J4} \end{eqnarray}
We see that due to annihilation of defects at collisions the total probability
diminishes at increasing the time separation $\tau$ as a power of $\tau$.
The property can be interpreted as follows: The majority of defect-antidefect
pairs annihilate fast after their creation and only a minor part of the defects
achieve a separation $r\gg{a}$. The probability of such event is proportional
to $(r/a)^{-2\beta}$.

The results obtained in the subsection are correct if the variation of
the coupling constant $\beta$ on the scale interval $a<r<\sqrt\tau$ is small.
The existence of such interval is justified by the assumed small value of
the fugacity $y_0$. Near $T_c$ variations of $\beta$ on a wide region of
scales can be relevant. Then the consideration needs a generalization made in
the last subsection of this section.

\subsection{Coulomb interaction and diffusion coefficient}

Let us consider the renormalization of the Coulomb interaction related to small
defect-antidefect pairs. It is known that the influence of such pairs can be
described in terms of a contribution to the effective dielectric constant, see
Ref. \cite{KT}. The picture is naturally generalized for the dynamics.

Before proceeding to calculations, it will be convenient for us to express
the Coulomb part (\ref{hamb}) of the Hamiltonian (\ref{ham}) in an
alternative form. Namely, using the Hubbard-Stratonovich trick we rewrite the
fourth-order term ${\cal H}_\beta$ as a functional integral over auxiliary
fields $\sigma$ and $\phi$
\begin{eqnarray} &&
\exp\left(-\int{\rm d}t\,{\cal H}_\beta\right)
\nonumber \\ &&
=\int{\cal D}\phi\,{\cal D}\sigma
\exp\left[-\int{\rm d}t\,
\left({\cal H}_1+{\cal H}_2\right)\right] \,,
\label{B0} \\ &&
{\cal H}_1=\int{\rm d}^2 r\,
\biggl[\left(\nabla\hat\psi_+\psi_+-\nabla\hat\psi_-\psi_-\right)\nabla\phi
\nonumber \\ &&
-\left(\hat\psi_+\psi_+-\hat\psi_-\psi_-\right)\sigma\biggr] \,,
\label{B1} \\ &&
{\cal H}_2=\frac{1}{4\pi\beta}\int{\rm d}^2 r\,\nabla\sigma\nabla\phi \,.
\label{B3} \end{eqnarray}
The relation (\ref{B0}) can be easily checked using the bare expression
\begin{eqnarray} &&
\langle\nabla\phi(t_1,{\bbox r}_1)
\sigma(t_2,{\bbox r}_2)\rangle_0=
-2\beta\frac{{\bbox r}_1-{\bbox r}_2}
{|{\bbox r}_1-{\bbox r}_2|^2} \,,
\label{B4} \end{eqnarray}
following from Eq. (\ref{B3}). We see that the correlation function (\ref{B4})
corresponds to the dashed line on the diagrams. Note that the field $\phi$
in the expressions is the electrostatic potential introduced by Eq.
(\ref{poten}). Indeed, integrating over the field $\sigma$ in Eq. (\ref{B0})
we get the Poisson equation
\begin{eqnarray} &&
\nabla^2\phi=-\frac{1}{4\pi\beta}\tilde\rho \,,
\label{BB4} \end{eqnarray}
leading to the expression (\ref{poten}).

Now we can work in terms of the sum
${\cal H}_0+{\cal H}_R+{\cal H}_1+{\cal H}_2$ where two first terms are
introduced by Eqs. (\ref{ham0},\ref{hamr}). Note that the sum
${\cal H}_0+{\cal H}_1$ is invariant under the following
infinitesimal transformation
\begin{eqnarray} &&
\delta\psi_+=\alpha\psi_+\, \quad
\delta\psi_-=-\alpha\psi_-\, \quad
\delta\hat\psi_+=-\alpha\psi_+
\nonumber \\ &&
\delta\hat\psi_-=\alpha\psi_-\, \quad
\delta\phi=2\alpha \, \quad
\delta\sigma=\nabla^2\alpha \,,
\label{ward}  \end{eqnarray}
where $\alpha$ is a function of coordinates. If to substitute
$R(r)\to\lambda\delta({\bbox r})$ and
$\bar R(r)\to\bar\lambda\delta({\bbox r})$ into the expression (\ref{hamr})
then it will be invariant under the transformation (\ref{ward}) also.
Deviations from the symmetry related to $r$-dependencies of $R$ and $\bar R$
are irrelevant. If the function $\alpha$ contains only terms linear and 
quadratic over ${\bbox r}$ then the contribution ${\cal H}_2$ (\ref{B3}) is
invariant under the transformation (\ref{ward}) also. The symmetry leads to
a number of the Ward identities. Particularly, they connect the renormalized
triple vertices to the self-energy function of the propagator.

\begin{figure}
\centerline{\psfig{figure=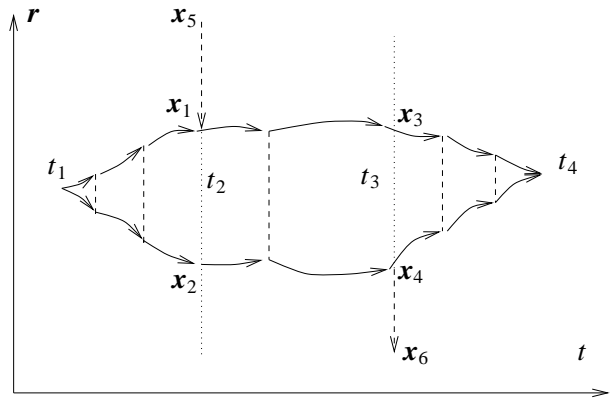,width=8cm}}
\caption{Typical diagram contributing to renormalization of
the effective ``dielectric constant''.}
\label{diag3}
\end{figure}

A typical diagram contributing to renormalization of the effective ``dielectric
constant'' is drawn in Fig. \ref{diag3}. There we see a loop composed of the
trajectories of a defect and of an antidefect which annihilate after their
creation. There are also two ``external'' dashed lines corresponding to the
interaction of the defect-antidefect pair with an environment. Besides the
diagrams of the type drawn in Fig. \ref{diag3} there are also diagrams with two
external dashed lines attached to the same trajectory. We draw the external
lines with arrows to remember that two sides of the dashed line representing
the correlation function (\ref{B4}) are not equivalent. We imply that the
dashed lines are directed from the field $\sigma$ to the field $\phi$.

As previously, we can dissect the diagram into parts which can be treated
separately. Then the answer can be found as a convolution of the corresponding
expressions. We perform the dissection along the planes in the ${\bbox r}-t$
space-time perpendicular to the $t$-axis and corresponding to the time moments
$t_2$ and $t_3$ of the external Coulomb lines. In Fig. \ref{diag3} the
dissection is shown by the dotted lines. We see that the loop is divided into
three parts.

The left part of the loop implying the integration over the time $t_1$ (see
Fig. \ref{diag3}) corresponds to
\begin{eqnarray} &&
\int_0^\infty{\rm d}\tau\, \int{\rm d}^2r_3\,{\rm d}^2r_4\,
\bar R({\bbox r}_3-{\bbox r}_4)
M(\tau,{\bbox x}_1,{\bbox x}_2,{\bbox r}_3,{\bbox r}_4)
\nonumber \\ &&
= \int_0^\infty{\rm d}\tau\,\bar R_\tau(|{\bbox x}_1-{\bbox x}_2|) \,.
\label{J6} \end{eqnarray}
Substituting here Eq. (\ref{fdt1}) and using Eq. (\ref{unity}) we get
\begin{eqnarray} &&
\int_0^\infty{\rm d}\tau\,\bar R_\tau(r)
=\frac{y_0^2}{a^4}
\left(\frac{a}{r}\right)^{2\beta} \,.
\label{J7} \end{eqnarray}
The central part of the diagram depicted in Fig. \ref{diag3} corresponds to the
conditional probability (\ref{mmm})
$M(t_3-t_2,{\bbox x}_3,{\bbox x}_4,{\bbox x}_1,{\bbox x}_2)$. And the
right part of the diagram in Fig. \ref{diag3} corresponds to the integral
(\ref{unity}). The relation can be recognized as a manifestation of
independence of all results of the final time $t_{\rm f}$ in the relation
(\ref{path1}). If we chose $t_{\rm f}=t_3$ then the right part of the
diagram in Fig. \ref{diag3} disappears and we should substitute $1$ instead,
in accordance with (\ref{unity}).

The structure of the diagram depicted in Fig. \ref{diag3} shows that the block
related to the defect-antidefect pair can be treated as a self-energy insertion
to the line corresponding to the Coulomb interaction. Thus it is natural to
expect that this insertion can be treated as a contribution to the ``dielectric
constant'', leading to a renormalization of the Coulomb constant $\beta$. The
corresponding quantitative analysis is presented in Appendix \ref{rebeta}
giving the following expression for the correction to $\beta$
\begin{eqnarray} &&
\Delta\beta\sim -y_0^2
\int\frac{{\rm d}r}{r}\left(\frac{a}{r}\right)^{2\beta-4} \,.
\label{B11} \end{eqnarray}
The expression can be treated as an integral over the characteristic
sizes of the defect-antidefect pairs.

\begin{figure}
\centerline{\psfig{figure=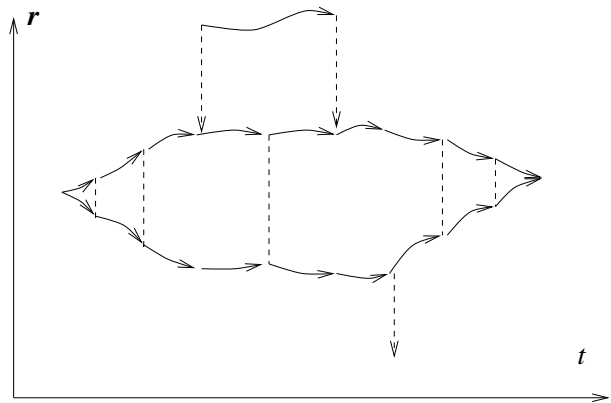,width=8cm}}
\caption{A more complicated diagram giving a correction to
the Coulomb interaction.}
\label{diag4}
\end{figure}

One may try to find more complicated blocks contributing to a renormalization
of the Coulomb coupling constant $\beta$. An example of such block is depicted
in Fig. \ref{diag4} where a number (three) ``external'' lines are attached to
the loop corresponding to the trajectories of the defect-antidefect pair.
One can easily check that the block depicted in Fig. \ref{diag4} gives a
correction to the Coulomb force which diminishes faster than $r^{-1}$ at
increasing the distance $r$ between the interacting particles. Therefore
the contribution is irrelevant. The same is true for more complicated
diagrams of the same type.

\begin{figure}
\centerline{\psfig{figure=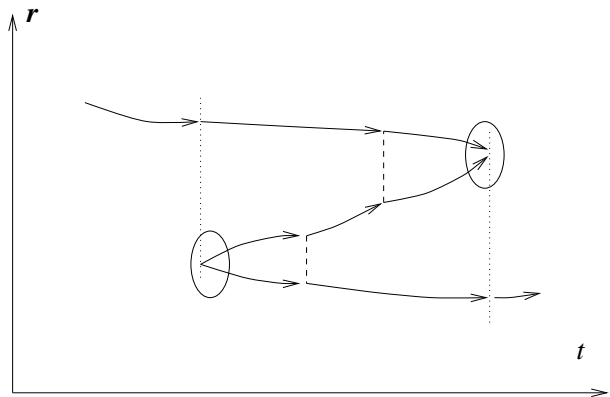,width=8cm}}
\caption{Illustration to the renormalization of
the diffusion coefficient.}
\label{diag5}
\end{figure}

We can also consider blocks which can be treated as contributions to the
diffusion coefficient $D$ introduced by Eqs. (\ref{H51},\ref{H52}). An
example is depicted in Fig. \ref{diag5}, where the block between two dotted
lines is a self-energy insertion to the propagator (\ref{diff}) which gives
the renormalization of the diffusion coefficient. We will assume that the
fields $\psi_\pm$ are corrected to keep the term (\ref{ham0}) unchanged. Then
the contribution (\ref{diff3}) has to be extracted from the renormalization
of the coefficient in front of the time derivatives.

To analyze the correction $\Delta D$ quantitatively one should know a
three-particle conditional probability which is more complicated than the
two-particle conditional probability (\ref{mmm}). Fortunately, one can estimate
the value of $\Delta D$ without detailed calculations. The point is that the
dependence of $\Delta D$ on the cutoff $a$ can be produced only by regions near
the creation or near the annihilation point (which are designated by ovals in
Fig. \ref{diag5}). The regions can be analyzed in terms of the two-particle
conditional probability (\ref{mmm}) since only the interaction of the nearest
``particles'' is relevant there. We already know the answer: the region near
the creation point produces the renormalized creation rate (\ref{J1})
whereas the region near the annihilation point produce no $a$-dependence. Then
simple dimensional estimates give the answer similar to the expression
(\ref{B11})
\begin{eqnarray} &&
\Delta D\sim -y_0^2
\int\frac{{\rm d}r}{r}\left(\frac{a}{r}\right)^{2\beta-4} \,.
\label{diff3} \end{eqnarray}
Remind that the bare value of $D$ is assumed to be equal to unity.

Of course there are also blocks which can be interpreted as corrections to the
triple vertices describing the interaction of the fields $\hat\psi_\pm$,
$\psi_\pm$ with the Coulomb fields $\phi$ and $\sigma$. An example of such
block can be imagined if to attach an external dashed line to a trajectory
between the dotted lines in Fig. \ref{diag5}. However, in the renormalization
scheme accepted the corrections are fully absorbed into the renormalization of
the fields $\psi_\pm$. That is a consequence of the symmetry of the
Hamiltonian under the transformation (\ref{ward}). Namely, the cubic term
(\ref{B1}) is unchanged provided the term (\ref{ham0}) is unchanged.

\subsection{Summary}

In the previous subsection we obtained the expressions for the corrections
(\ref{B11},\ref{diff3}) of the Coulomb constant and of the diffusion
coefficient in the main order over the fugacity $y_0$. Now we are going to
discuss high-order corrections over $y_0$ which in statics lead to the
renormalization group equations (\ref{rg}). It will be more convenient for us 
to proceed in spirit of the Kosterlitz renormalization group scheme. 
Namely, we see that the expressions 
(\ref{B11},\ref{diff3}) are written as integrals over the
space variable $r$ which can be treated as the size of a defect-antidefect
pair. We can first perform the integration over a restricted interval of the
sizes, what gives slightly renormalized values of the coupling constants. Then 
we can repeat the integration. In the limit this multi-step procedure gives 
the renormalization group equations for the coupling constants, as Kosterlitz
suggested. On the diagrammatic language the procedure means that we gradually
substitute blocks corresponding to small separations between the particles by
their effective values relative to larger scales. The procedure can also be
considered as increasing an effective size of the defects $a\to r$. Then the
renormalization of the coupling constants can be described in terms of the
differential renormalization group equations.

At each step of the procedure we deal with correlation functions like
(\ref{mmm}). For an interval of scales where a variation of the Coulomb
constant $\beta$ is small one can use for the function the expression
(\ref{H70}) where now one should substitute the renormalized value of the
Coulomb constant $\beta$. Analogously, the renormalized annihilation rate is
determined by Eq. (\ref{dres1}) where one should substitute the expression
(\ref{H69}) with the renormalized value of the Coulomb constant $\beta$.
Next, for the renormalized creation rate we should use the relation
\begin{eqnarray} &&
\bar R_{\tau_1}=\int{\rm d}^2 r_0\,
S(\tau_1-\tau_2,{\bbox r},{\bbox r}_0)
\bar R_{\tau_2} \,,
\label{JJ1} \end{eqnarray}
where $\tau_1>\tau_2$. The relation (\ref{JJ1}) can be derived from Eq.
(\ref{J1}) if to use the property of $S$ analogous to Eq. (\ref{conv}). For
the renormalized creation rate the relation (\ref{JJ1}) is correct only if
$\beta(r_1)$ weakly differs from $\beta(r_2)$ for characteristic values of
the parameters $r_1\sim\sqrt{\tau_1}$ and $r_2\sim\sqrt{\tau_2}$.

A relation analogous to Eq. (\ref{JJ1}) can be formulated for the
annihilation rate $R$. The relation leads to the same expression
(\ref{dres1}) where $\beta$ is now scale-dependent. Therefore the
renormalized quantity of the annihilation constant $\lambda_{\rm r}$ flows
together with $\beta$ in accordance with Eq. (\ref{rrr}). That is accounted
for the non-logarithmic character of the integrals leading to the relation
(\ref{rrr}).

Then we should define the renormalized fugacity in dynamics. For the purpose
let us generalize the expression (\ref{J7})
\begin{eqnarray} &&
\int_0^\infty{\rm d}\tau\,\bar R_\tau(r)
=\frac{y^2}{r^4}\,.
\label{JJ7} \end{eqnarray}
Then Eq. (\ref{J7}) is rewritten as
\begin{eqnarray} &&
y=\left(\frac{a}{r}\right)^{\beta-2}y_0 \,.
\label{renoy} \end{eqnarray}
The relation can be treated as an elementary step of the renormalization group
procedure which is described by Eq. (\ref{rg} for $y$. Thus the renormalization
group equation for $\beta$ in dynamics coincides with one in statics.

The expression (\ref{B11}) for the correction to the Coulomb constant $\beta$
can be considered as arising at the elementary step of the renormalization 
group procedure. The corresponding renormalization group equation 
can be found if to 
pass to the differential form and to substitute $y_0$ by the renormalized 
value $y$ in accordance with Eq. (\ref{renoy}). Then we obtain the 
renormalization group equation coinciding with Eq. (\ref{rg}) for $\beta$. 
The expression (\ref{diff3}) for the correction to the diffusion coefficient 
leads to the following renormalization group equation
\begin{eqnarray} &&
\frac{{\rm d}D}{{\rm d}\ln(r/a)}\sim -y^2 \,,
\label{diff4} \end{eqnarray}
analogous to the equation (\ref{rg}) for $\beta$. We conclude that the
correction to $D$ is small due to the small value of the fugacity and is
therefore irrelevant. To avoid a misunderstanding, remind that the variation of
the Coulomb constant $\beta$ with increasing scale is also small. Nevertheless,
as seen from Eq. (\ref{rg}), it is the difference $\beta-2$ that enters the
renormalization group equations and the variation of the difference can be 
essential.

\section{Correlation Functions}
\label{corr}

Here, we treat non-simultaneous correlation functions of the charge density
$\rho$ (\ref{den})
\begin{eqnarray} &&
F_{2n}(t_1,\dots,t_{2n};{\bbox r}_1,\dots,{\bbox r}_{2n})
\nonumber \\ &&
=\left\langle \rho(t_1,{\bbox r}_1) \dots
\rho(t_{2n},{\bbox r}_{2n})\right\rangle \,.
\label{H30} \end{eqnarray}
Note an obvious consequence of the constraint (\ref{Ch1})
\begin{eqnarray} &&
\int{\rm d}^2 r_1\, F_{2n}=0 \,.
\label{Ch2} \end{eqnarray}
To examine the correlation functions (\ref{H30}) we use the representation
(\ref{do3}). We will assume that all the diagrammatic blocks corresponding to
small defect-antidefect pairs are already included into the renormalization
of the corresponding coupling constants as discussed in Section \ref{renorm}.
Therefore the fugacity $y$ and the Coulomb constant $\beta$ entering all
subsequent expressions should be taken at the current scale. First we will
examine contributions to the functions associated with a single
defect-antidefect pair and then we will consider contributions related to a
number of defect-antidefect pairs.

\subsection{Pair Correlation Function}

We start with the pair correlation function
\begin{eqnarray} &&
F_2(t_2-t_1,{\bbox r}_2-{\bbox r}_1)
=\langle\rho(t_2,{\bbox r}_2)
\rho(t_1,{\bbox r}_1)\rangle \,,
\label{J22} \end{eqnarray}
with $t_2>t_1$. The average (\ref{J22}) can be calculated in accordance
with the relation (\ref{path1}) where one should substitute the expression
(\ref{do3}). We assume $t_{\rm f}=t_2$.

\begin{figure}
\centerline{\psfig{figure=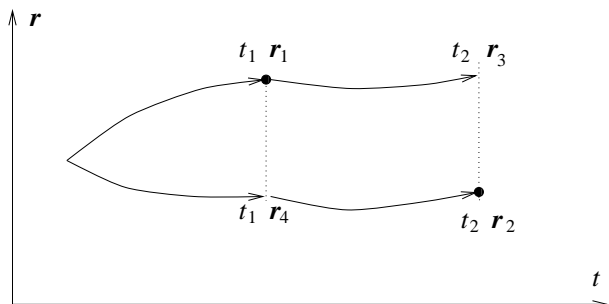,width=8cm}}
\caption{Trajectories contributing to
the pair correlation function.}
\label{vort4}
\end{figure}

The contribution to the average (\ref{J22}) related to a single
defect-antidefect pair can be represented by a series of the diagrams with
two lines constituted of the defect and antidefect propagators. The lines
start from the creation point. A half of the diagrams have the structure
depicted in Fig. \ref{vort4}. Here we omitted lines and vertices
corresponding to the interaction of the defects (which is implied) and keep
only trajectories of the defects. The trajectories should pass through the
points ${\bbox r}_1$ and ${\bbox r}_2$ at the time moments $t_1$ and $t_2$
(the events are designated by black circles). An additional contribution to
the average (\ref{J22}) is determined by similar diagrams where both events
$t_1,{\bbox r}_1$ and $t_2,{\bbox r}_2$ belong to the same trajectory.

As above, we dissect the diagram into parts which can be treated separately.
Let us make a cut along planes in ${\bbox r}-t$ space-time corresponding to 
the time moments $t_1$ and $t_2$, they are shown in Fig. \ref{vort4} by dotted
lines. Intermediate points appearing in the convolution (\ref{conv}) are
designated in Fig. \ref{vort4} as ${\bbox r}_3$ and ${\bbox r}_4$. After that
the diagram is divided into two parts separated by the dotted line. The part of
the diagram to the right from the dotted line corresponds to the conditional
probability $M$ (\ref{mmm}) and the part of the diagram to the left from the
dotted line corresponds to the correlation function
\begin{eqnarray} &&
\Phi_2({\bbox r}_1-{\bbox r}_2)
=\langle \psi_+(t,{\bbox r}_1)\psi_-(t,{\bbox r}_2)\rangle \,.
\label{J21} \end{eqnarray}

The quantity (\ref{J21}) can be treated as the probability density to find a
defect-antidefect pair with a given separation. Correspondingly, the integral
$\int{\rm d}^2r\,\Phi_2({\bbox r})$ determines the density of the
defect-antidefect pairs. The correlation function (\ref{J21}) coincides with
the integral in the left-hand side of Eq. (\ref{JJ7}). Hence
\begin{eqnarray} &&
\Phi_2(r)=y^2/r^4  \,,
\label{rela} \end{eqnarray}
where $y$ is the renormalized fugacity. The equations (\ref{fug},\ref{rela})
show that asymptotically in the low-temperature phase
\begin{eqnarray} &&
\Phi_2(r)\propto r^{-2\beta}\,.
\label{new} \end{eqnarray}
Note that the same behavior (\ref{new}) is observed up to a slowly varying
factor in the whole region of scales.

The diagram depicted in Fig. \ref{vort4} gives a convolution of $\Phi_2$ and
$M$. Adding the contribution corresponding to the case where both events
$t_1,{\bbox r}_1$ and $t_2,{\bbox r}_2$ belong to the same trajectory we get
the following expression for the pair correlation function (\ref{J22})
\begin{eqnarray} &&
F_2(t,{\bbox r}_2-{\bbox r}_1)=
-2\int{\rm d}^2 r_3\,{\rm d}^2 r_4\,
\Phi_2({\bbox r}_4-{\bbox r}_1)
\nonumber \\ &&
\times \left[M(t,{\bbox r}_3,{\bbox r}_2,{\bbox r}_1,{\bbox r}_4)
-M(t,{\bbox r}_2,{\bbox r}_3,{\bbox r}_1,{\bbox r}_4)\right] \,,
\label{H81} \end{eqnarray}
where $t>0$. At small times $t$ we turn to the limit law (\ref{small}).
Substituting the expression into Eq. (\ref{H81}) we get
\begin{eqnarray} &&
F_2(t=0,{\bbox r}_1-{\bbox r}_2)
\nonumber \\ &&
=2\delta({\bbox r}_1-{\bbox r}_2)\int{\rm d}^2r\,\Phi_2({\bbox r})
-2\Phi_2({\bbox r}_1-{\bbox r}_2) \,.
\label{H83} \end{eqnarray}
Here the second contribution corresponds to the law (\ref{pair}) and the term
proportional to $\delta$-function is an autocorrelation contribution associated
with a single defect. The factor in front of the $\delta$-function (which is
the density of defects) is in accordance with the relation (\ref{Ch2}). Note
that at small $t$ the $\delta$-function is converted into a narrow function of
the width $\sim\sqrt t$.

It follows from Eqs. (\ref{H70},\ref{rela},\ref{H81}) that for $t\sim r^2$
\begin{eqnarray} &&
F_2(t,{\bbox r})\sim\frac{y^2(r)}{r^4} \,,
\label{HH83} \end{eqnarray}
To justify Eq. (\ref{HH83}) one should check that there are no divergences in
the integral (\ref{H81}). It can be done directly using Eq. (\ref{H70}). The
convergence at small separations $r$ and $r_0$ is accounted for the behavior of
the modified Bessel functions $I_\nu(x)\propto x^\nu$ at small values of the
argument. The convergence at large separations $r$ and $r_0$ can be checked
using the asymptotic law (\ref{W8}). If $|t|\gg r^2$ then
\begin{eqnarray} &&
F_2\sim -\frac{y^2(r)}{r^{4-2\beta}|t|^\beta} \,.
\label{H84} \end{eqnarray}
The asymptotic law is established in Appendix \ref{pairco}.

The behavior of the pair correlation function determined by the laws
(\ref{HH83},\ref{H84}) corresponds to a conventional critical dynamics (see,
e.g., Ref. \cite{HH}) with the dynamical critical index $z=2$. However, as we
will see below, the behavior of the high-order correlation functions is beyond
the conventional scheme. Besides, the scaling law $t\sim r^2$ is true for the
high-order correlation functions as well.

\subsection{High-Order Correlation functions}
\label{high-o}

Here we extend the procedure of the preceding subsection to the case of the
high-order correlation functions $F_{2n}$ (\ref{H30}). We will assume that
$t_1<t_2<\dots<t_{2n}$. Again, we examine the contribution to $F_{2n}$
associated with a single defect-antidefect pair. Corresponding diagrams contain
two trajectories starting anywhere and passing through the points
${\bbox r}_1$, \dots, ${\bbox r}_{2n}$ at the time moments $t_1$, \dots,
$t_{2n}$. We will designate the trajectories of the defect and of the
antidefect as ${\bbox x}(t)$ and ${\bbox z}(t)$.

\begin{figure}
\centerline{\psfig{figure=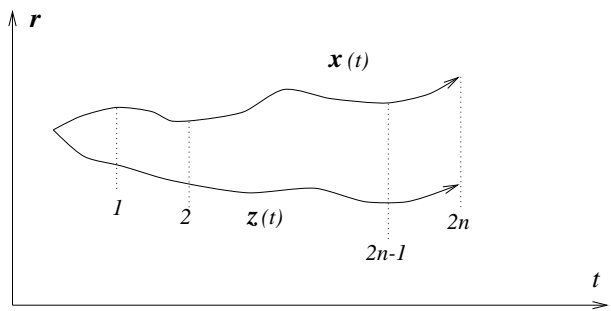,width=8cm}}
\caption{Trajectories passing through a number of points.}
\label{vort5}
\end{figure}

Let us dissect the diagrams along planes in the ${\bbox r}-t$ space-time
corresponding to the time moments $t_1$, \dots, $t_{2n}$. Then the diagram is
divided into a number of blocks, see Fig. \ref{vort5}. The left block in Fig.
\ref{vort5} corresponds to the object (\ref{J21}) and all the other blocks
correspond to the correlation function (\ref{mmm}).  Again, using Eq.
(\ref{conv}) we can write the contribution to the correlation function
(\ref{H30}) associated with a single defect-antidefect pair as the following
convolution
\begin{eqnarray} &&
F_{2n}(t_1,\dots,t_{2n};{\bbox r}_1,\dots{\bbox r}_{2n})
\nonumber \\ &&
=\prod\limits_{j=1}^{2n}\int{\rm d}^2 x_j\,{\rm d}^2z_j\,
\Phi_{2}({\bbox x}_1-{\bbox z}_1)
\left[\delta({\bbox r}_j-{\bbox x}_j)
-\delta({\bbox r}_j-{\bbox z}_j)\right]
\nonumber \\ &&
\times M(t_{j+1}-t_j,{\bbox x}_{j+1},{\bbox z}_{j+1},
{\bbox x}_{j},{\bbox z}_{j})\,.
\label{J31} \end{eqnarray}
Here, one must replace the last factor $M(t_{2n+1}-t_{2n})$ by unity. The
relation (\ref{J31}) is a generalization of Eq. (\ref{H81}). Thus we got an
expression for the correlation function which is a multiple integral of
functions determined by explicit formulas. A recurrent procedure for
calculating $F_{2n}$ is suggested in Appendix \ref{recurr}.

Of course the expression (\ref{H81}) for the pair correlation function is
reproduced by Eq. (\ref{J31}). Note also the following expression

\end{multicols}

\begin{eqnarray} &&
\langle\rho(t,{\bbox r}_1)\rho(t,{\bbox r}_2)
\rho(0,{\bbox r}_3)\rho(0,{\bbox r}_4)\rangle
=2\left[M(t,{\bbox r}_1,{\bbox r}_2,{\bbox r}_3,{\bbox r}_4)
+M(t,{\bbox r}_1,{\bbox r}_2,{\bbox r}_4,{\bbox r}_3)\right]
\Phi_2({\bbox r}_3-{\bbox r}_4)
\nonumber \\ &&
-B({\bbox r}_1-{\bbox r}_2)\delta({\bbox r}_3-{\bbox r}_4)
-B({\bbox r}_3-{\bbox r}_4)\delta({\bbox r}_1-{\bbox r}_2)
+\int{\rm d}^2r\,B({\bbox r})\,
\delta({\bbox r}_1-{\bbox r}_2)
\delta({\bbox r}_3-{\bbox r}_4) \,,
\label{fou} \end{eqnarray}

\begin{multicols}{2}

\noindent
where
\begin{eqnarray} &&
B({\bbox r}_3-{\bbox r}_4)=2\int{\rm d}^2r_1\,
\bigl[M(t,{\bbox r}_1,{\bbox r}_2,{\bbox r}_3,{\bbox r}_4)
\nonumber \\ &&
+M(t,{\bbox r}_1,{\bbox r}_2,{\bbox r}_4,{\bbox r}_3)\bigr]
\Phi_2({\bbox r}_3-{\bbox r}_4) \,.
\nonumber \end{eqnarray}
The formula can be found from Eq. (\ref{J31}) using the relation (\ref{small}).
Naturally, the expression (\ref{fou}) is symmetric under the permutation
${\bbox r}_1,{\bbox r}_2\leftrightarrow{\bbox r}_3,{\bbox r}_4$, which can be
checked using the expressions (\ref{H70},\ref{new}). The formula (\ref{fou}) is
in agreement with the general property (\ref{Ch2}).

The recurrent procedure for calculating $F_{2n}$ suggested in Appendix
\ref{recurr} shows that there are no divergences in the integrals at all steps
of calculating $F_{2n}$. This means that we can evaluate the correlation
functions from naive dimension estimates. Namely, if all space separations
among $|{\bbox r}_i-{\bbox r}_j|$ are of the same order $r_*$ and all time
intervals are of the order $r_*^2$ then
\begin{eqnarray} &&
F_{2n}\sim y^2(r_*)r_*^{-4n} \,.
\label{H91} \end{eqnarray}
In the large-scale limit when $\beta$ is saturated we have in accordance with
Eq. (\ref{fug})
\begin{eqnarray} &&
F_{2n}\propto r_*^{-4(n-1)-2\beta} \,.
\label{H99} \end{eqnarray}

If some space separations among $|{\bbox r}_i-{\bbox r}_j|$ and/or some time
intervals differ strongly then one can formulate some simple rules following
from Eqs. (\ref{H70},\ref{J31}). Let us give some examples. If one of the time
intervals $\tau$ is much larger than all values of the squared separations
$|{\bbox r}_i-{\bbox r}_j|^2$ then the correlation function behaves like
$F_{2n}\propto\tau^{-\beta}$. It can be proved like it is done in Appendix
\ref{pairco} for the pair correlation function. For small $\tau$ there appear
contributions to $F_{2n}$ short-correlated in space (on scales
$\sim\sqrt\tau$). In the limit $\tau\to0$ the contributions turn into
$\delta$-functions, as it was for the pair correlation function, see Eq.
(\ref{H83}), representing an autocorrelation of single defects. If the points
${\bbox r}_j$ can be divided into two ``clouds'' with a separation $r$ between
the clouds much larger than their sizes (and all time intervals are much
smaller than $r^2$) then the principal $r$-dependence of the correlation
function $F_{2n}$ is the same as in the function $\Phi_2(r)$ (\ref{rela}).

Remember that the charge density $\rho$ is related to the curl of the gradient
of the hexatic angle $\varphi$ for hexatics and to the curl of the gradient
of the order parameter phase for superfluid films. It is instructive to
re-express our results in terms of the phase gradient circulations of
$\nabla\varphi$ over a closed loop $C$
\begin{eqnarray} &&
\Gamma(t,C)=\oint{\rm d}{\bbox r}\,\nabla\varphi
=2\pi\int {\rm d}^2 r\,\rho(t,{\bbox r}) \,,
\label{circ} \end{eqnarray}
where the second integral is taken over the area inside the loop.
Correlation functions of $\Gamma$'s can be rewritten as integrals of the
correlation functions $F_{2n}$. As an example, consider the following average
\begin{eqnarray} &&
\Psi_{2n}=\bigl\langle \Gamma(t,C)\Gamma(t+\tau,C)\dots
\nonumber \\ &&
\times \Gamma[t+(2n-1)\tau,C]\bigr\rangle  \,.
\label{aver} \end{eqnarray}
Suppose that the characteristic size of the loop $r$ is large enough so
that we can assume that $\beta$ is saturated, and that $\tau\sim r^2$. Then
the following scaling law is satisfied: if $r\to Xr$ and $\tau\to X^2 \tau$
then
\begin{eqnarray} &&
\Psi_{2n}\to X^{4-2\beta}\Psi_{2n} \,,
\label{scal} \end{eqnarray}
where $X$ is an arbitrary factor. The law (\ref{scal}) is a consequence of
Eq. (\ref{H99}). It has two striking peculiarities. First, it possesses a
clear critical dependence. Second, it is independent of the order $n$.

The procedure described above can be generalized to include the external
potential. Then one should take the solution of the equation (\ref{MM1}) for
the function (\ref{mmm}) and the corresponding expression for the object
(\ref{J21}).

\subsection{Many-Pair Contributions}

\begin{figure}
\centerline{\psfig{figure=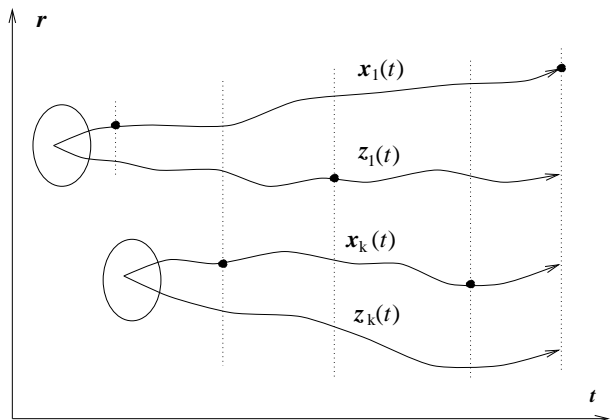,width=8cm}}
\caption{A number of defect-antidefect pairs passing through given points.}
\label{vort6}
\end{figure}

We have established the contributions to the charge density correlation
functions $F_{2n}$ associated with a single defect-antidefect pair. Now we are
going to discuss other contributions to the correlation functions related to an
arbitrary number of defect-antidefect pairs. Correspondingly, we should take
diagrams with a number of trajectories passing through the points
${\bbox r}_1$, \dots, ${\bbox r}_{2n}$ at the time moments $t_1$, \dots,
$t_{2n}$. The picture illustrating the situation is drawn in Fig. \ref{vort6}
where black circles correspond to the arguments of $F_{2n}$:
($t_1,{\bbox r}_1$), \dots, ($t_{2n},{\bbox r}_{2n}$). There we omitted
blocks related to short-living defect-antidefect pairs regarding that the
blocks are already included into the renormalization of the Coulomb constant
$\beta$.

As previously, we can dissect the diagrams along the planes in the
${\bbox r}-t$ space-time corresponding to the time moments
$t_1$, \dots, $t_{2n}$. Then any diagram will be divided into a number of
strips, see Fig.  \ref{vort6}. The part of the diagram within each strip can
be treated as the corresponding matrix element of the evolution operator
$\exp(-\int{\rm d}t\,{\cal H})$. Then the contribution to $F_{2n}$
will be written like (\ref{J31}) as a convolution of the matrix elements.
Generally, the matrix elements can be estimated like the function
(\ref{mmm}): each trajectory segment gives a factor which scales as $1/t$ and
$t$ scales as $r^2$. But there are obvious exceptions from the rule.  Namely,
going back in time we will come to a moment where a defect-antidefect pair
was created.  Again, when we consider small separations between the defect
and the antidefect (which is correct for time moments close to the creation
time) we can take into account only the Coulomb interaction between the two
created defects. The corresponding regions in Fig. \ref{vort6} are inside the
ovals.  Each such region produces the factor $y^2$. Therefore generally
$F_{2n}\propto y^{2k}$, where $k$ is the number of the pairs. Taking into
account also a scale-dependent factor we get
\begin{eqnarray} &&
F_{2n}\sim y^{2k}(r_*)r_*^{-4n} \,,
\label{J41} \end{eqnarray}
where we assume that all space separations are of the order $r_*$ and all
time intervals are of the order $r_*^2$.

The expression (\ref{J41}) is a generalization of Eq. (\ref{H91}). Comparing
these two expressions we see that the ratio of the contribution (\ref{J41}) to
the contribution (\ref{H91}) is the $(k-1)$-th power of a dimensionless small
parameter $y^2(r_*)$. Thus we conclude that the leading contribution to
$F_{2n}$ is related to a single defect-antidefect pair, that corresponds to
$k=1$.  Now we can explain the origin of the estimate (\ref{simul1}) for the
simultaneous correlation functions which obviously does not coincide with Eq.
(\ref{H91}). The estimate (\ref{simul1}) in terms of Eq. (\ref{J41})
corresponds to $k=n$. The reason is quite obvious: Two defects cannot pass
simultaneously through $2n$ points and at least $k=n$ defect-antidefect pairs
should be taken to obtain a non-zero contribution to the simultaneous
correlation function $F_{2n}$. The situation is illustrated by Fig.
\ref{vort7}. Note that the estimate (\ref{simul1}) is not correct for the
autocorrelation contributions proportional to $\delta$-functions, as written in
Eq. (\ref{H83}).

Thus we have two different regimes: for simultaneous and for non-simultaneous
correlation functions. Let us establish the boundary between the regimes. For
the purpose we should consider small time intervals where the single-pair
contribution is finite but small. The smallness is associated with diffusive
exponents presented, e.g., in the expression (\ref{H70}). Therefore the
characteristic time where the simultaneous regime passes into the
non-simultaneous one can be estimated as
\begin{eqnarray} &&
t\sim\frac{r^2}{|\ln[y(r)]|} \,,
\label{simul} \end{eqnarray}
where $r$ is a space separation corresponding to the small time interval.
In the low-temperature phase on large scales (where $\beta$ is saturated)
we have $|\ln y|\approx(\beta -2)\ln(r/a)$.

\section{Superfluid films}
\label{helium}

Let us consider superfluid films. The equation of motion for the vortices
contains an additional term (Magnus force). Thus instead of Eq. (\ref{H51})
we should write (see Ref. \cite{Dynamics1})
\begin{eqnarray} &&
\frac{{\rm d}x_{j,\alpha}}{{\rm d}t}=-\frac{D}{T}
\left[\frac{\partial{\cal F}}{\partial x_{\alpha j}}
+n_j\gamma\epsilon_{\alpha\beta}
\frac{\partial{\cal F}}{\partial x_{\beta j}}\right]
+{\xi}_{j,\alpha} \,,
\label{H93} \end{eqnarray}
where $\gamma$ is a new dimensionless parameter. The equation (\ref{H93})
can be derived in the spirit of the procedure proposed by Hall and Vinen for
the $3d$ superfluid, see Ref. \cite{HV}. Huber \cite{82Hu} argued that the
same equation is correct for spin vortices in planar $2d$ magnetics.

For the superfluid films the ``charge density'' (\ref{den}) is proportional to
the vorticity ${\rm curl}\,{\bbox v}_{\rm s}$. To calculate correlation
functions $F_{2n}$ (\ref{H30}) one can use the scheme developed in the previous
sections. The only difference is that instead of the expression (\ref{H70})
for the conditional probability $M$ (\ref{mmm}) one should use the solution of
the equation
\begin{eqnarray} &&
\partial_t M=\left(\frac{1}{2}\nabla_\varrho^2+
2\nabla_r^2+4\beta\frac{\bbox r}{r^2}\nabla_r\right)M
+2\gamma\beta\epsilon_{\alpha\beta}\frac{r_\beta}{r^2}
\nabla_{\varrho\alpha}M
\nonumber \\ &&
-R(r)M+\delta(t)\delta({\bbox r}-{\bbox r}_0)
\delta\left({\bbox\varrho}-\frac{{\bbox r}_3}{2}
-\frac{{\bbox r}_4}{2}\right) \,.
\label{H95} \end{eqnarray}
The variables ${\bbox r}$, ${\bbox r}_0$ and ${\bbox\varrho}$ are introduced
by Eq. (\ref{J3}). Again, on scales $r\gg a$ one can omit the term with the
annihilation rate $R$ in Eq. (\ref{H95}) demanding a finite value of $M$ at
$r\to 0$ instead.

Unfortunately, a cross term over ${\bbox r}$ and ${\bbox\varrho}$ appears
in the operator in the right-hand side of Eq. (\ref{H95}). Thus one cannot
obtain an explicit expression for $M$ of the type of Eq. (\ref{H70}).
Nevertheless, this additional term has the same dimensionality as the other
terms and does not change the scaling estimates $M\sim t^{-2}$, $t\sim r^2$
determining the function $M$. Moreover, the equation for the object
\begin{eqnarray} &&
S(t,{\bbox r},{\bbox r}_0)=\int{\rm d}^2\varrho\,
M(t,{\bbox r},{\bbox\varrho},{\bbox r}_0,{\bbox r}_3/2+{\bbox r}_4/2)  \,,
\label{H96} \end{eqnarray}
following from (\ref{H95}) is identical to Eq. (\ref{four}). Therefore for
the object (\ref{H96}) we have the same series (\ref{angle}) with the
coefficients (\ref{H69}).

Looking through the derivation presented in Section \ref{renorm} we see that
just the function $S$ (\ref{H96}) enters all the relations. Therefore we can
make the same assertions as previously. First, on large scales the annihilation
coefficient $\lambda$ is equal to its universal value (\ref{rrr}). Second, we
can write the same expression (\ref{rela}) for the average (\ref{J21}). Third,
in dynamics we get the same renormalization group equation (\ref{rg}) for 
$\beta$, see Appendix \ref{rebeta}. Fourth, the renormalization of the 
diffusion coefficient is irrelevant. And finally, one can assert that a 
renormalization of the parameter $\gamma$ introduced by Eq. (\ref{H93}) 
is determined by the equation
\begin{eqnarray} &&
\frac{{\rm d}\gamma}{{\rm d}\ln(r/a)}\sim y^2 \,,
\nonumber \end{eqnarray}
analogous to (\ref{diff4}). Therefore the renormalization of $\gamma$ is
irrelevant. Again, the scheme can be generalized to include the ``external
potential'', which now is the average value of the superfluid velocity.

Next, we proceed to the correlation functions $F_{2n}$ (\ref{H30}).
Formally they are determined by the same convolution (\ref{J31}) as previously.
However, one should substitute there the solution of the equation (\ref{H95}).
Therefore the concrete expressions for $F_{2n}$ will be different.
Nevertheless, the estimates like (\ref{H91},\ref{H99},\ref{J41}) remains true
because of the following reasons. First, due to the same dimensionality of all
the terms in the right-hand side of Eq. (\ref{H95}) the function $M$ possesses
the simple scaling properties noted above. Second, there are no divergences in
the convolutions like (\ref{J31}) determining the objects. To prove the second
property, we should analyze a behavior of $M$ at large and at small
separations. In the case  $rr_0/t\gg1$ the characteristic values of the
separations ${\bbox r}_1-{\bbox r}_3$ and of ${\bbox r}_2-{\bbox r}_4$ are
$\sim\sqrt t$ and are consequently much smaller than
$|{\bbox r}_1-{\bbox r}_2|$ (or $|{\bbox r}_3-{\bbox r}_4|$). Then it is
possible to neglect all terms containing $|{\bbox r}_1-{\bbox r}_2|$ in
denominators in the equation (\ref{H95}) and we come to a purely diffusive
equation leading to the asymptotics (\ref{W8}). It is possible to establish
that the small-scale of the conditional probability $M$ for the vortices
coincides with that examined above. The properties ensure convergence of all
intermediate integrals appearing at calculating the correlation functions of
vorticity $F_{2n}$.

We have also the same scaling law (\ref{scal}) for the correlation function
(\ref{aver}) of the integrals (\ref{circ}) which are now proportional to the
circulations of the superfluid velocity. We conclude that all the scaling laws
for the correlation functions of the vorticity and their asymptotic behavior
remains the same as previously.

\section{Discussion}
\label{conclu}

The main result of our consideration is the expression (\ref{H91}) for
high-order correlation functions of the ``charge density'' (\ref{H30}) which is
disclinicity for hexatic films and vorticity for superfluid films. We see from
Eq. (\ref{H91}) that the high-order correlation functions are much larger than
their normal estimates via the pair correlation function. Namely, in accordance
with Eqs. (\ref{HH83},\ref{H91}) we have
\begin{eqnarray} &&
{F_{2n}}/{F_2^n}\sim y^{-2n+2}\gg1 \,,
\label{intem} \end{eqnarray}
where $y$ is the renormalized fugacity. The asymptotic behavior of the ratio
at large scales is determined by the law (\ref{fug}). Though at developing
our scheme we accepted that the defect-antidefect pairs constitute a dilute
solution we hope that the scaling law (\ref{intem}) is universal. The ground
for the hope is the renormalization group procedure 
(formulated in Ref. \cite{74Kos}) 
which shows that on large scales we come to an effectively dilute solution of 
the pairs. We believe that the most interesting fact to be compared with
experiment or numerics is the scaling law (\ref{scal}) which is a
consequence of Eq. (\ref{intem}).

The physics behind the inequality (\ref{intem}) is as follows. The main
contribution to the correlation functions is associated with a single
defect-antidefect pair. Though the contribution associated with a number of
defect-antidefect pairs contains an additional huge entropy factor it has
also an additional small factor associated with small probability to observe
defect-antidefect pairs with separations larger than the core radius $a$. The
smallness is accounted for the strong Coulomb attraction. The considered
effect is a consequence of the competition of those two factors. The result
of the competition manifests in the law (\ref{J41}) which gives the estimate
for the contribution associated with $k$ defect-antidefect pairs. For
simultaneous correlation functions nothing similar occurs and we have the
conventional estimate (\ref{simul1}). That is the reason why the effect
cannot be observed in statics. The property is directly related to causality
since a defect-antidefect pair cannot simultaneously pass through $2n$ points
and at least $n$ defect-antidefect pairs is needed to get a non-zero
contribution to the simultaneous correlation function $F_{2n}$, see Fig.
\ref{vort7}. That explains the estimate (\ref{simul1}). Thus we have two
different regimes for simultaneous and non-simultaneous correlation
functions. The characteristic boundary time separating those two regimes is
written in Eq. (\ref{simul}).

The considered effect resembles intermittency in turbulence (see, e.g., Ref.
\cite{Frish}) leading to large $r$-dependent factors in the ratios like
(\ref{intem}) in the velocity correlation functions of a turbulent flow.
However, as is seen from Eq. (\ref{intem}), for the defects the large
$r$-dependent factors are related to the ultraviolet cutoff parameter ${a}$
whereas for intermittency in turbulence the large $r$-dependent factors are
related to the infrared (pumping) scale. Our situation is thus closer to the
inverse cascade (see Ref. \cite{Kraich}) realized on scales much larger than
the pumping length. There are experimental data \cite{98Tab} concerning the
inverse cascade in $2d$ hydrodynamics and analytical observations concerning
the inverse cascade for a compressible fluid \cite{CKGV} which indicate the
absence of the intermittency in the inverse cascades.  Note that only
simultaneous objects were examined in the works \cite{98Tab,CKGV}, and there is
no intermittency in our simultaneous correlation functions. So, based on the
analogy, one may think that for the inverse cascades non-simultaneous objects
reveal some intermittency.

The consideration presented in our work is applicable to superfluid films.
There exist also films and quasi-$2d$ systems of different symmetry. Huber
\cite{82Hu} argued that the same equation as for quantum vortices is correct
for spin vortices in planar $2d$ magnets. We believe that our approach based
on the equation (\ref{H51}) is correct for the dynamics of disclinations in
hexatic films like membranes, freely suspended films and Langmuir films.
Next, the above scheme seems to work also for dislocations in solid films.
The system needs a special treatment since a modification should be
introduced into the procedure. Maybe some features of the presented picture
can be observed also in superconductive materials, especially in high-$T_c$
superconductors. There are analytical and numerical indications that for a
purely Langevin dynamics of the order parameter there are logarithmic
corrections to the law (\ref{H51}), see Refs.
\cite{87Cl,88Pl,91Pa,91Ry,92Mu,93Yu,94Ko}. We believe that the logarithms are
destroyed if to switch on an interaction of the order parameter with other
degrees of freedom. Nevertheless, it would be interesting to generalize our
scheme including the logarithmic corrections.

\acknowledgements

I am grateful to G. Falkovich, K. Gawedzki, I. Kolokolov, and M. Vergassola
for useful discussions and to E. Balkovsky, A. Kashuba and S. Korshunov for
valuable remarks. This research was supported in part by a grant of Israel
Science Foundation, by a grant of Minerva Foundation and by the
Landau-Weizmann Prize program.

\appendix

\section{}

In this Appendix we present some calculations leading to the results
presented in the main body.

\subsection{Some relations for the pair conditional probability}
\label{condi}

Here, we deduce some relations concerning the correlation function
(\ref{mmm}). We examine the solution of the equations (\ref{four}) with the
annihilation term $R$ omitted. Then we should accept a finite value of $S_m$
at $r\to 0$ as is explained in the main text.

Performing the Fourier transform over $t$ we get from Eq. (\ref{four})
\begin{eqnarray} &&
\frac{-i\omega}{2}S_m(\omega)
=\left[\partial_r^2+(1+2\beta)\frac{1}{r}\partial_r
-\frac{m^2}{r^2}\right]S_m(\omega)
\nonumber \\ &&
+\frac{1}{4\pi r_0}\delta(r-r_0) \,.
\label{H67} \end{eqnarray}
The solution of the equation (\ref{H67}) which tends to zero at $r\to\infty$
and remains finite at $r\to0$ can be written as
\begin{eqnarray} &&
S_m(\omega)=\frac{1}{4\pi}\left(\frac{r_0}{r}\right)^\beta
I_\nu\left(\sqrt{\frac{-i\omega}{2}}\,r\right)
\nonumber \\ &&
\times K_\nu\left(\sqrt{\frac{-i\omega}{2}}\,r_0\right)
\qquad {\rm if} \ r<r_0 \,,
\nonumber \\ &&
S_m(\omega)=\frac{1}{4\pi}\left(\frac{r_0}{r}\right)^\beta
K_\nu\left(\sqrt{\frac{-i\omega}{2}}\,r\right)
\nonumber \\ &&
\times I_\nu\left(\sqrt{\frac{-i\omega}{2}}\,r_0\right)
\qquad {\rm if} \ r>r_0 \,,
\label{H68} \end{eqnarray}
where $\nu=\sqrt{\beta^2+m^2}$ and we used the relation
\begin{eqnarray} &&
K_\nu(z)\partial_z I_\nu(z)
-I_\nu(z)\partial_z K_\nu(z)=z^{-1} \,.
\nonumber \end{eqnarray}
At the next step one performs the inverse Fourier transform. Deforming
the integration contour to the negative imaginary semi-axis and using the
relation 6.633.2 from Ref. \cite{GR} one gets Eq. (\ref{H69}).

Next, we present two integral relations for the conditional probability
(\ref{mmm}) in the above approximation. Let us calculate the total
probability to find a defect and an antidefect in any points at a fixed time
separation $t$. Using the relation 6.631.1 from Ref. \cite{GR} we get
\begin{eqnarray} &&
\int{\rm d}^2r_1\,{\rm d}^2r_2\,M
=\int{\rm d}^2r\,S(t,{\bbox r},{\bbox r}_0)
\label{dres2} \\ &&
=\frac{1}{\Gamma(1+\beta)}
\left(\frac{r_0^2}{8t}\right)^\beta
\exp\left(-\frac{r_0^2}{8t}\right)
~_1\!F_1\left(1,\beta+1;\frac{r_0^2}{8t}\right) \,.
\nonumber \end{eqnarray}
We conclude from Eq. (\ref{dres2}) that the total probability diminishes with
increasing time $t$. It is accounted for annihilation of defects at collisions.
Using the relation 6.611.4 written in Ref. \cite{GR} we find
from (\ref{H70})
\begin{eqnarray} &&
\int\limits_0^\infty {\rm d} t\, M =
\frac{1}{2\pi^2}\left(\frac{r_0}{r}\right)^\beta
\sum\limits_{m=-\infty}^{+\infty}\exp(im\varphi)\
\nonumber \\ &&
\times \frac{(2rr_0)^\nu}{\sqrt{\varsigma^2-4r^2r_0^2}
\left(\varsigma+\sqrt{\varsigma^2-4r^2r_0^2}\right)^\nu} \,,
\label{H71}\end{eqnarray}
where $\varsigma=\left({\bbox r}_1+{\bbox r}_2-{\bbox r}_3
-{\bbox r}_4\right)^2+r^2+r_0^2$ and $\nu=\sqrt{\beta^2+m^2}$.
The expression determines a distribution of the defect and of the
antidefect over space provided the pair was created near the origin
at any time.

Now we calculate the integral (\ref{dres3}). Substituting there
Eqs. (\ref{dres1},\ref{dres2}) we get after integrating in part
\begin{eqnarray} &&
\lambda_{\rm r}=-8\pi\int\limits_0^\infty{\rm d}z\,
\left[\frac{z^\beta e^{-z}}{\Gamma(1+\beta)}
~_1\!F_1(1,\beta+1;z)-1\right] \,,
\label{dre3} \end{eqnarray}
where $z=r_0^2/8\tau$. One can easily check using the asymptotic expression
for $~_1\!F_1(1,\beta+1;z)$ at large $z$ that the integral (\ref{dre3})
converges. Let us now rewrite the expression (\ref{dre3}) as
\begin{eqnarray} &&
\lambda_{\rm r}=-8\pi\lim_{\alpha\to0}
\int\limits_0^\infty{\rm d}z\,
\biggl[\frac{z^\beta}{\Gamma(1+\beta)}\exp(-z-\alpha z)
\nonumber \\ &&
\times_1\!F_1(1,\beta+1;z)
-\exp(-\alpha z)\biggr] \,.
\label{dre4} \end{eqnarray}
Now integrals of both contributions to the integrand can be found explicitly
(see the relation 7.621.5 from Ref. \cite{GR}). Passing then to the
limit $\alpha\to0$ we come to the answer (\ref{rrr}).

\subsection{Asymptotics of the Pair Conditional Probability}
\label{asympt}

Here, we will be interested in the asymptotic behavior of the expression
(\ref{H70}) where the parameter $z=rr_0/(4t)$ is large: $z\gg1$.
More precisely, we will examine the function
\begin{eqnarray} &&
T(z,\varphi)=\sum\limits_{m=-\infty}^{+\infty}\exp(im\varphi)\
I_\nu\left(z\right) \,,
\label{W1} \end{eqnarray}
where $\nu=\sqrt{\beta^2+m^2}$. It is obvious that
$T(z,\varphi)=T(z,-\varphi)$. Below we will assume $\varphi>0$.

First of all, we convert the sum over $m$ in Eq. (\ref{W1}) into an integral
using the identity
\begin{eqnarray} &&
\sum\limits_{m=-\infty}^{+\infty} f(m)=
\frac{1}{2}\int_+{\rm d}m\, \coth(-i\pi m) f(m)
\nonumber \\ &&
-\frac{1}{2}\int_-{\rm d}m\, \coth(-i\pi m) f(m) \,,
\nonumber \end{eqnarray}
where the contour in the first integral goes above the real axis and
the contour in the second integral goes below the real axis. One can
easily check that if $f(-m)$ is complex conjugated to $f(m)$ then the
second integral is equal to the complex conjugated of the first integral
with the sign minus. Therefore we can write
\begin{eqnarray} &&
T(z,\varphi)={\rm Re}\int_+{\rm d}m\, \coth(-i\pi m)
\exp(im\varphi)\,I_\nu\left(z\right) \,.
\label{W2} \end{eqnarray}
Next, we are going to shift the integration contour in Eq. (\ref{W2})
into the upper $m$-semi-plane. Then we should separately consider two
contributions to $I_\nu$ which are determined by two terms in its integral
representation
\begin{eqnarray} &&
I_\nu(x)=\frac{1}{\pi}\int_{0}^\pi{\rm d}\vartheta\,
\exp(x\cos\vartheta)\cos(\nu\vartheta) \,.
\label{A08} \end{eqnarray}
Substituting the first contribution in the right-hand side of Eq. (\ref{A08})
we get
\begin{eqnarray} &&
T_1(z,\varphi)={\rm Re}\int_+{\rm d}m\, \coth(-i\pi m)
\nonumber \\ &&
\times \int_{0}^\pi\frac{{\rm d}\vartheta}{\pi}
\exp(z\cos\vartheta)\exp(im\varphi)\cos(\nu\vartheta) \,.
\label{W3} \end{eqnarray}
As we will see, the answer will be determined by a vicinity of a saddle point
where $-im\gg1$. There one can substitute $\coth(-i\pi m)$ by unity and $\nu$
by $m$. Then one obtains from Eq. (\ref{W3}) the following saddle-point
conditions
\begin{eqnarray} &&
\vartheta=\varphi\,, \quad
m=iz\sin\vartheta \,.
\nonumber \end{eqnarray}
One can easily find from Eq. (\ref{W3}) in the saddle-point approximation
\begin{eqnarray} &&
T_1(z,\varphi)\approx\exp(z\cos\varphi) \,,
\label{W5} \end{eqnarray}
where we have taken into account an pre-exponent besides the exponent.
The above scheme is obviously broken at small $\varphi$ or at $\varphi$
close to $\pi$ since the saddle-point value of $m$ tends to zero there.
Thus the cases need a separate consideration.

Let us consider the case where $\varphi$ is close to zero. Then the main
contribution to Eq. (\ref{W3}) is gained from the region of integration over
$\vartheta$ near zero. There it is possible to expand the factor at $z$ as
$\cos\vartheta\approx1-\vartheta^2/2$. Then the integration over $\varphi$
is performed explicitly. Returning also to the sum we get
\begin{eqnarray} &&
T_1(z,\varphi)\approx\frac{\exp z}{\sqrt{2\pi z}}
\sum\limits_{m=-\infty}^{+\infty}
\exp\left(im\varphi-\frac{m^2}{2z}\right) \,,
\nonumber  \end{eqnarray}
where we substituted $\exp(-\beta^2/2z)$ by unity.
Since both $\varphi\ll1$ and $1/z\ll1$ we can replace here summation by the
integration over $m$. After the substitution we get
\begin{eqnarray} &&
T_1(z,\varphi)\approx
\exp\left(z-z\varphi^2/2\right) \,,
\nonumber  \end{eqnarray}
what reproduces Eq. (\ref{W5}) at small $\varphi$. Analogously the case of
$\varphi$ close to $\pi$ can be examined. Again, we reproduce Eq. (\ref{W5}).

Next, we examine the contribution to the function $T$ associated with the
second term in (\ref{A08}). For the purpose we return to the initial expression
(\ref{W1}). The second contribution can be written as
\begin{eqnarray} &&
T_2(z,\varphi)=-\frac{1}{\pi}\int_0^\infty{\rm d}t\,
\sum\limits_{m=-\infty}^{+\infty}
\exp(im\varphi-\nu t)\
\nonumber \\ &&
\times \sin(\pi\nu)\exp(-z\cosh t) \,.
\nonumber \end{eqnarray}
If the sum is dominated by first terms in the sum over $m$ then we have
\begin{eqnarray} &&
|T_2(z,\varphi)|\lesssim\frac{1}{\sqrt z}\exp(-z) \,,
\label{W6} \end{eqnarray}
which can be obtained after substituting $\cosh t\approx 1+t^2/2$ into
the integral. Thus we should estimate the contribution determined by
large $|m|$. Then
\begin{eqnarray} &&
\sin(\pi\nu)\approx(-1)^m\frac{\pi\beta^2}{2|m|} \,.
\nonumber \end{eqnarray}
Substituting the expression into the sum we get a contribution to
$T_2(z,\varphi)$
\begin{eqnarray} &&
-\int_0^\infty{\rm d}t\,\exp(-z\cosh t)
\nonumber \\ &&
\times \sum\limits_{m\gg1}\frac{\beta^2}{|m|}
\cos[(\pi-\varphi)m]\exp(-m t)\
\nonumber \\ &&
=\int_0^\infty{\rm d}t\,\exp(-z\cosh t)\
\Phi\left[{t^2+(\pi-\varphi)^2}\right] \,,
\nonumber \end{eqnarray}
where we introduced a function $\Phi(x)$ which behaves as $\Phi(x)\sim\ln x$ if
$x\ll1$ and $|\Phi|\ll1$ if $x\gg1$. Therefore the contribution determined by
large $|m|$ reproduces the same estimate (\ref{W6}). The estimate (\ref{W6})
shows that $|T_2|$ is much smaller than Eq. (\ref{W5}). We conclude that the
expression (\ref{W5}) determines the main contribution to the sum (\ref{W1}).

Rewriting the answer (\ref{W5}) for the sum (\ref{W1})
in terms of the original variables we get
\begin{eqnarray} &&
\sum\limits_{m=-\infty}^{+\infty}\exp(im\varphi)\
I_\nu\left(\frac{rr_0}{4t}\right)
\approx\exp\left(\frac{{\bbox r}{\bbox r}_0}{4t}\right) \,.
\label{W7} \end{eqnarray}
Then the expression (\ref{W7}) leads to
\begin{eqnarray} &&
M\approx\frac{1}{(4\pi t)^2}\left(\frac{r_0}{r}\right)^\beta
\exp\left\{-\frac{({\bbox r}-{\bbox r}_0)^2}{8t}\right\}
\nonumber \\ &&
\times \exp\left\{-\frac{\left({\bbox r}_1+{\bbox r}_2-{\bbox r}_3
-{\bbox r}_4\right)^2}{8t}\right\} \,.
\label{WW8} \end{eqnarray}

\subsection{Asymptotics of the pair correlation function}
\label{pairco}

Let us consider the asymptotics of $F_2$ for times $t$ much larger than
$r_{12}^2$. Then the characteristic values of both ${r}$ and ${r}_0$ in the
integral (\ref{H81}) are much larger than $r_{12}$ and therefore we can neglect
a dependence on $r_{12}$ there. Substituting the expressions
(\ref{H70},\ref{J7}) into Eq. (\ref{H81}) and then neglecting $r_{12}$ we get
after integrating over the angle between ${\bbox r}$ and ${\bbox r}_0$
\begin{eqnarray} &&
F_2\approx-\frac{y^2}{r^{4-2\beta}t^2}
\int\frac{{\rm d} r}{r^{\beta-1}}
\frac{{\rm d} r_0}{r_0^{\beta-1}}
\exp\left(-\frac{r^2+r_0^2}{4t}\right)
\nonumber \\ &&
\times \sum\limits_{m=-\infty}^{\infty}
I_{2m+1}\left(\frac{rr_0}{4t}\right)
I_{\nu}\left(\frac{rr_0}{4t}\right) \,,
\nonumber \end{eqnarray}
where $\nu=\sqrt{\beta^2+(2m+1)^2}$ and ${\bbox r}_0={\bbox r}_1-{\bbox r}_4$,
${\bbox r}={\bbox r}_3-{\bbox r}_2$. The ratio $y^2/r^{4-2\beta}$ is a slowly
varying function and therefore it is placed outside the integral.
Note that the integral is nonzero due
to $I_\nu>0$. Performing here the integration over $r/r_0$ we get
\begin{eqnarray} &&
F_2\approx-\frac{y^2}{r^{4-2\beta}t^2}
\int_0^\infty\frac{{\rm d}y}{y^{\beta-1}}K_0(2y)
\nonumber \\ &&
\sum\limits_{m=-\infty}^{\infty}
I_{2m+1}\left(y\right)I_{\nu}\left(y\right) \,,
\nonumber \end{eqnarray}
where we used the relation 3.547.4 from Ref. \cite{GR}. The integral over $y$
converges at any $m$ obviously. Thus we should examine the convergence of the
sum over $m$. At large $\nu$ we can use the asymptotics
\begin{eqnarray} &&
I_\nu(x)\approx\frac{1}{\sqrt{2\pi}}
\frac{1}{(x^2+\nu^2)^{1/4}}
\nonumber \\ &&
\times \exp\left[\sqrt{x^2+\nu^2}+\nu\ln\frac{x}
{\nu+\sqrt{x^2+\nu^2}}\right] \,,
\label{J07} \end{eqnarray}
and the integral over $y$ is gained at large $y$, where the asymptotics
$K_0(2y)\approx\sqrt{\pi/4y}\,\exp(-2y)$ works. Therefore
\begin{eqnarray} &&
\int_0^\infty\frac{{\rm d}y}{y^{\beta-1}}K_0(2y)
I_{2m+1}\left(y\right)I_{\nu}\left(y\right)
\nonumber \\ &&
\sim\frac{1}{\nu^{\beta-1/2}}
\int_0^\infty\frac{{\rm d}z}{z^{\beta-1/2}\sqrt{z^2+1}}
\exp(2\nu f) \,,
\nonumber \\ &&
f(z)=\sqrt{z^2+1}\,-z
+\ln\frac{z}{1+\sqrt{z^2+1}}\,.
\nonumber \end{eqnarray}
Since the function $f(z)$ increases monotonically, the integral over $z$ is
gained where $f\sim\nu^{-1}$ that is at $z\sim\nu$. That leads to an estimate
\begin{eqnarray} &&
\int_0^\infty\frac{{\rm d}y}{y^{\beta-1}}K_0(2y)
I_{2m+1}\left(y\right)I_{\nu}\left(y\right)
\sim\frac{1}{m^{2\beta-1}} \,.
\nonumber \end{eqnarray}
The estimate means that the sum over $m$ converges at large $m$ since
$\beta>2$. Thus we come to the estimate (\ref{H84}).

\section{Correction to the Coulomb coupling constant}
\label{rebeta}

Here, we calculate the correction to the Coulomb coupling constant $\beta$
produced by small defect-antidefect pairs. We start from Eqs.
(\ref{B1},\ref{B3}). The correction can be found from the relation
\begin{eqnarray} &&
\exp\left[-\int{\rm d}t\,\Delta{\cal H}_2(\phi,\sigma)\right]
\label{B5} \\ &&
=\biggl\langle\exp\biggl\{\int {\rm d}t\,{\rm d}^2 r\,
\biggl[\left(-\nabla\hat\psi_+\psi_++\nabla\hat\psi_-\psi_-\right)\nabla\phi
\nonumber \\ &&
+\left(\hat\psi_+\psi_+-\hat\psi_-\psi_-\right)\sigma\biggl]\biggr\}
\biggr\rangle \,,
\nonumber\end{eqnarray}
where the angular brackets mean averaging over the small defect-antidefect
pairs and $\phi$, $\sigma$ in the exponent are to be treated as ``external''
fields. Really, one of the contributions to $\Delta{\cal H}$ is depicted in
Fig. \ref{diag3} where the upper arrowed dashed line correspond to the
field $\phi$ and the lower arrowed dashed line correspond to the field $\sigma$.
The other contribution to $\Delta{\cal H}$ is associated with a similar
diagram where both ``external'' dashed line are attached to the same trajectory.
It is explained in Sec. \ref{renorm} what objects correspond to the different
parts of the diagram. Here we present an analytical expression written in
accordance with the two noted diagrams (see designations of the points in
Fig. \ref{diag3})
\begin{eqnarray} &&
\Delta{\cal H}_2=2 \int {\rm d}t_2\,{\rm d}t_3\,
{\rm d}^2 x_1\,{\rm d}^2 x_2\,{\rm d}^2 x_3\,{\rm d}^2 x_4
\nonumber \\ &&
\times \nabla_\alpha\phi(t_2,{\bbox x}_1)
[\sigma(t_3,{\bbox x}_4)-\sigma(t_3,{\bbox x}_3)]
\nonumber \\ &&
\times \nabla_{1\alpha}\Phi_2({\bbox x}_1-{\bbox x}_2)
M(t_3-t_2,{\bbox x}_3,{\bbox x}_4,{\bbox x}_1,{\bbox x}_2) \,,
\label{B6} \end{eqnarray}
where the objects $\Phi_2$ and $M$ are defined by Eqs. (\ref{mmm},\ref{J21}).
The explicit expressions for the functions are written in Eqs.
(\ref{H70},\ref{J7}).

We are interested in the situation when the ``external'' fields $\phi$ and
$\sigma$ in Eq. (\ref{B6}) vary on scales much larger than a characteristic
size of the defect-antidefect pair. Then one can substitute
\begin{eqnarray} &&
\sigma(t_3,{\bbox x}_4)-\sigma(t_3,{\bbox x}_3)\to
\nabla_\gamma \sigma(t_2,{\bbox x}_1)
({\bbox x}_{4\gamma}-{\bbox x}_{3\gamma}) \,,
\nonumber \end{eqnarray}
where we expanded the difference and then shift the argument of the factor
$\nabla\sigma$. Substituting the expression into Eq. (\ref{B6}) and
integrating over ${\bbox\varrho}=({\bbox x}_3+{\bbox x}_4)/2$ we get
\begin{eqnarray} &&
\Delta{\cal H}_2=-2 \int {\rm d}t\,{\rm d}^2 r\,
\nabla_\alpha\phi(t,{\bbox r})
\nabla_\gamma \sigma(t,{\bbox r})
\label{B7} \\ &&
\times \int {\rm d}\tau\,{\rm d}^2 x\,{\rm d}^2 r_0\,
x_\gamma\nabla_\alpha\Phi_2({\bbox r}_0)
S(\tau,{\bbox x},{\bbox r}_0) \,,
\nonumber \end{eqnarray}
where ${\bbox r}_0={\bbox x}_1-{\bbox x}_2$,
${\bbox x}={\bbox x}_3-{\bbox x}_4$, $t=t_2$, $\tau=t_3-t_2$,
${\bbox r}={\bbox x}_1$,  and we substituted the
expression (\ref{H65}). Next, the integral (\ref{B7})
is proportional to $\delta_{\alpha\beta}$ (due to averaging over angles).
Taking into account also the expressions (\ref{angle},\ref{J7}) we get
\begin{eqnarray} &&
\Delta{\cal H}_2=4\pi^2\beta\int {\rm d}t\,{\rm d}^2 r\,
\nabla\phi(t,{\bbox r})
\nabla \sigma(t,{\bbox r})
\nonumber \\ &&
\times \int\limits_0^\infty {\rm d}\tau\,{\rm d}x\,{\rm d}r_0\,
x^2\Phi_2({r}_0)S(\tau,x,{r}_0) \,.
\label{B8} \end{eqnarray}
Comparing the expression with Eq. (\ref{B3}) we conclude that the integral
in Eq. (\ref{B8}) can be considered as a correction to the coefficient
$1/(4\pi\beta)$. Thus we get
\begin{eqnarray} &&
\Delta\beta=-16\pi^3\beta^3
\int\limits_0^\infty {\rm d}\tau\,{\rm d}x\,{\rm d}r_0\,
x^2\Phi_2({r}_0)S(\tau,x,{r}_0) \,.
\label{B9} \end{eqnarray}
Substituting here the expression (\ref{H69}) we obtain
\begin{eqnarray} &&
\Delta\beta=-\frac{2\pi^2\beta^3}{\nu}
\int{\rm d}x\,{\rm d}r_0\,x^2\Phi_2({r}_0)
\nonumber \\ &&
\times \frac{\left(x^2+r_0^2-|x^2-r_0^2|\right)^{\nu}}{(2xr_0)^{\nu}}
\left(\frac{r_0}{x}\right)^\beta
\label{B10} \end{eqnarray}
where $\nu=\sqrt{\beta^2+1}$ and we used the relation 6.623.3 from Ref.
\cite{GR}. Now, remembering the expression (\ref{J7}), one can easily
check that the integral over $r_0$ in Eq. (\ref{B10}) converges both for
small and large $r_0$. Therefore the integral is gained at $r_0\sim x$.
Thus, substituting Eq. (\ref{J7}), we get from Eq. (\ref{B10}) the answer
(\ref{B11}).

Now some words about generalization to the case of superfluid films discussed
in Section \ref{helium}. The new term in the right-hand side of Eq. (\ref{H93})
produces an additional contribution to the ``Hamiltonian'' of the system. Then
the term (\ref{B1}) should be substituted by
\begin{eqnarray} &&
{\cal H}_1=-\int{\rm d}^2 r\,
\biggl[\left(\hat\psi_+\nabla_\alpha\psi_+
-\hat\psi_-\nabla_\alpha\psi_-\right)
\nabla_\alpha\phi
\label{B20} \\ &&
+\left(\hat\psi_+\nabla_\alpha\psi_+
+\hat\psi_-\nabla_\alpha\psi_-\right)
\gamma\epsilon_{\alpha\beta}\nabla_\beta\phi
\nonumber \\ &&
+\left(\hat\psi_+\psi_+
-\hat\psi_-\psi_-\right)\sigma\biggr] \,.
\nonumber \end{eqnarray}
The expression (\ref{B20}) means a modification of the triple vertices shown by
the arrows in diagrams, see e.g. Fig. \ref{diag3}. Next, we can repeat the same
steps as above starting from the Hamiltonian determined by Eq. (\ref{B20}).
Then the intermediate expressions like Eqs. (\ref{B6},\ref{B7}) will be
modified. The terms which does not contain $\gamma$ are reduced to the
conditional probability $S$ and give the same correction (\ref{B10}) to
$\beta$. The terms containing $\gamma$ do not produce relevant terms at all
because of the identity
\begin{eqnarray} &&
\int{\rm d}^2 r\,\nabla_\alpha\sigma
\epsilon_{\alpha\beta}\nabla_\beta\phi=0 \,.
\nonumber  \end{eqnarray}
Thus for the superfluid films we get the same answers (\ref{B10},\ref{B11}).

\section{Recurrent procedure}
\label{recurr}

We see from the expression (\ref{J31}) that the correlation function $F_{2n}$
can be calculated step by step: first integrating over
${\bbox x}_{1},{\bbox z}_{1}$, then over ${\bbox x}_2,{\bbox z}_2$ and so
further. To formulate the procedure one should introduce auxiliary objects
\begin{eqnarray} &&
H_k(t_1,\dots,t_k;{\bbox r}_1,\dots,{\bbox r}_{k-1};
{\bbox x}_{k},{\bbox z}_{k})
\nonumber \\ &&
=\prod\limits_{j=1}^{k-1}\int{\rm d}^2 x_j\,{\rm d}^2z_j\,
\left[\delta({\bbox r}_j-{\bbox x}_j)
-\delta({\bbox r}_j-{\bbox z}_j)\right]
\nonumber \\ &&
\times M(t_{k}-t_{k-1},{\bbox x}_{k},{\bbox z}_{k},
{\bbox x}_{k-1},{\bbox z}_{k-1})\dots
\nonumber \\ &&
\times M(t_{2}-t_{1},{\bbox x}_{2},{\bbox z}_{2},
{\bbox x}_{1},{\bbox z}_{1})\Phi_2({\bbox x}_{1}-{\bbox z}_{1}) \,.
\label{J32} \end{eqnarray}
One can easily check that the functions $H_k$ are symmetric under permuting
${\bbox x}_{k}$ and ${\bbox z}_{k}$ for odd $k$ and are antisymmetric under the
permutation for even $k$. That depends on the number of factors containing
differences of $\delta$-functions which are antisymmetric under permuting
${\bbox x}_j$ and ${\bbox z}_j$ whereas the factors $M$ are symmetric under the
permutation. The correlation function $F_{2n}$ can be restored from $H_{2n}$:
\begin{eqnarray} &&
F_{2n}=2\int{\rm d}^2\varrho\,
H_{2n}(t_i;{\bbox r}_1,\dots,{\bbox r}_{2n-1},
{\bbox r}_{2n};{\bbox\varrho}) \,,
\label{H89} \end{eqnarray}
as follows from Eq. (\ref{J31}). Note that an attempt to calculate
odd correlation functions $F_{2n+1}$ from $H_{2n+1}$ using the same scheme
gives zero (as it should be) since $H_{2n+1}$ is symmetric under permuting
${\bbox x}_{2n+1}$ and ${\bbox z}_{2n+1}$ whereas the difference
$\delta({\bbox r}-{\bbox x}_{2n+1})-\delta({\bbox r}-{\bbox z}_{2n+1})$ is
antisymmetric under the permutation.

To obtain $H_k$ we can use a recurrent scheme. First of all we conclude
from Eq. (\ref{J32}) that
\begin{eqnarray} &&
H_1(t,{\bbox x},{\bbox z})=\Phi_2({\bbox x}-{\bbox z}) \,.
\label{JJ32} \end{eqnarray}
Next, one can easily obtain from the definition (\ref{J32}) the
following recurrent relation
\begin{eqnarray} &&
H_{k+1}(t_1,\dots,t_{k+1};{\bbox r}_1,\dots,{\bbox r}_{k},
{\bbox x}_{k+1},{\bbox z}_{k+1})
\nonumber \\ &&
=\int{\rm d}^2 x_k\,{\rm d}^2z_k\,
\left[\delta({\bbox r}_k-{\bbox x}_k)
-\delta({\bbox r}_k-{\bbox z}_k)\right]
\nonumber \\ &&
\times M(t_{k+1}-t_{k},{\bbox x}_{k+1},{\bbox z}_{k+1},
{\bbox x}_{k},{\bbox z}_{k})
\nonumber \\ &&
\times H_k(t_1,\dots,t_k;{\bbox r}_1,\dots,{\bbox r}_{k-1},
{\bbox x}_{k},{\bbox z}_{k}) \,.
\label{J33} \end{eqnarray}
Taking into account the symmetry properties of $H_k$ we can rewrite the
relation (\ref{J33}) as

\end{multicols}

\begin{eqnarray} &&
H_{2n}(t_1,\dots,t_{2n};{\bbox r}_1,\dots,{\bbox r}_{2n-1},
{\bbox x}_{2n},{\bbox z}_{2n})
=-\int{\rm d}^2\varrho\,H_{2n-1}(t_1,\dots,t_{2n-1};{\bbox r}_1,\dots,
{\bbox r}_{2n-1},{\bbox r}_{2n},{\bbox\varrho})
\nonumber \\ &&
\times\left[M(t_{2n}-t_{2n-1},{\bbox x}_{2n},{\bbox z}_{2n},
{\bbox\varrho},{\bbox r}_{2n-1})
-M(t_{2n}-t_{2n-1},{\bbox x}_{2n},{\bbox z}_{2n},
{\bbox r}_{2n-1},{\bbox\varrho}) \right] \,,
\label{H87} \\ &&
H_{2n+1}(t_1,\dots,t_{2n+1};{\bbox r}_1,\dots,{\bbox r}_{2n},
{\bbox x}_{2n+1},{\bbox z}_{2n+1})
=\int{\rm d}^2\varrho\,H_{2n}(t_1,\dots,t_{2n};{\bbox r}_1,\dots,
{\bbox r}_{2n},{\bbox r}_{2n+1},{\bbox\varrho})
\nonumber \\ &&
\times\left[M(t_{2n+1}-t_{2n},{\bbox x}_{2n+1},{\bbox z}_{2n+1},
{\bbox\varrho},{\bbox r}_{2n})
+M(t_{2n+1}-t_{2n},{\bbox x}_{2n+1},{\bbox z}_{2n+1},
{\bbox r}_{2n},{\bbox\varrho}) \right] \,.
\label{H88} \end{eqnarray}

\begin{multicols}{2}

\noindent
Thus we can subsequently obtain the functions $H_k$ and then get
the correlation functions $F_{2n}$ in accordance with Eq. (\ref{H89}).

Now we are going to establish scaling properties of the correlation functions.
As is seen from the recurrent relations (\ref{H87},\ref{H88}) and Eq.
(\ref{H89}) the crucial point is convergence of integrals over ${\bbox\varrho}$
there. There are no problems with finite $\rho$ since one can conclude from Eq.
(\ref{H70}) that $M$ remains finite at any values of its space arguments.
Indeed, it follows from Eq. (\ref{H70}) that the only dangerous case is
$r\to0$. Then the arguments of $I_\nu$ also tend to zero. Substituting there
the first terms of the expansion $I_\nu(x)\approx(x/2)^\nu/\Gamma(1+\nu)$ we
conclude that the value of $M$ remains finite at $r\to0$. Therefore we should
examine a behavior of the integrands in Eqs. (\ref{H87},\ref{H88}) at large
${\bbox\varrho}$. The behavior is dominated by the asymptotics (\ref{W8}).
First of all note that at large $\varrho$
\begin{eqnarray} &&
H_k(t_1,\dots,t_k;{\bbox r}_1,\dots,{\bbox r}_{k-1},{\bbox x}_{k},
{\bbox\varrho})\propto\varrho^{-2\beta} \,.
\label{H90} \end{eqnarray}
The behavior (\ref{H90}) is characteristic of $\Phi_2$, that is characteristic
of $H_1$, see Eqs. (\ref{JJ32},\ref{J7}). Then the behavior is reproduced at
each step of the recurrent procedure (\ref{H87},\ref{H88}). Indeed, in the
first terms in Eqs.  (\ref{H87},\ref{H88}) $M\sim1$ if ${\bbox\varrho}$ is
close to ${\bbox x}$ and $M$ exponentially decays at increasing
${\bbox\varrho}-{\bbox x}$. Therefore at large ${\bbox x}$ the integrals
(\ref{H87},\ref{H88}) over ${\bbox\varrho}$ are gained near ${\bbox x}$, that
leads to the above assertion. The asymptotic behavior (\ref{H90}) ensures
convergence of the integral (\ref{H89}) determining the correlation functions.

Thus there are no divergences in the integrals at all steps of calculating
$F_{2n}$ in accordance with (\ref{H87},\ref{H88}) and (\ref{H89}).

\end{multicols}

\end{document}